\documentclass[journal]{IEEEtran}
\usepackage{textcomp}

\usepackage{amsmath}
\usepackage{graphicx}
\usepackage{epstopdf}
\usepackage{subfigure}
\usepackage{amsmath}
\usepackage{algorithm}
\usepackage{algpseudocode}
\usepackage{cite}
\usepackage{multicol}
\usepackage{multirow}
\usepackage{mathtools}
\usepackage{amssymb}
\usepackage{amsthm}
\usepackage{color}
\usepackage{bm}
\usepackage{indentfirst}
\usepackage{url}
\usepackage{nomencl}
\usepackage{booktabs}
\usepackage{cite}
\usepackage[numbers,sort&compress]{natbib}
\usepackage{amsmath}
\usepackage{graphicx}
\usepackage{subfigure}
\usepackage{amsmath}
\usepackage{algorithm}
\usepackage{algpseudocode}
\usepackage{cite}
\definecolor{mygray}{gray}{.9}
\usepackage{multicol}
\usepackage{multirow}
\usepackage{mathtools}
\usepackage{amssymb}
\usepackage{color}
\usepackage{bm}
\usepackage{indentfirst}
\usepackage{url}
\usepackage{nomencl}
\usepackage{booktabs}
\usepackage{cite}
\usepackage[numbers,sort&compress]{natbib}
\newcommand{\tabincell}[2]{\begin{tabular}{@{}#1@{}}#2\end{tabular}}
\newtheorem{Theorem}{Theorem}
\newtheorem{Lemma}{Lemma}

\newtheorem{Remark}{Remark}
\newtheorem{Assumption}{Assumption}
\allowdisplaybreaks

\begin{document}

\title{Social Cost Optimization for Prosumer Community with Two Price-Package Incentives in Two-Settlement Based Electricity Market}%
%
%

\author{Jianzheng Wang and Guoqiang Hu
\thanks{Jianzheng Wang and Guoqiang Hu are with the School of Electrical and Electronic Engineering, Nanyang Technological University, Singapore 639798 (email:
        {\tt\small wang1151@e.ntu.edu.sg, gqhu@ntu.edu.sg}).}%
}
\maketitle
\thispagestyle{empty}
\pagestyle{empty}

\begin{abstract}
In this paper, we consider a future electricity market consisting of aggregated energy prosumers, who are equipped with local {{wind power plants}} (WPPs) to support (part of) their energy demands and can also trade energy with {{day-ahead market}} (DAM) and {{energy balancing market}} (EBM). In addition, an {{energy aggregator}} (EA) is established, who can provide the trading gateways between prosumers and the markets. The EA is responsible for making pricing strategies on the prosumers to influence their trading behaviours such that the social benefit of the prosumer community is improved. Specifically, two price packages are provided by the EA: {{wholesale price}} (WP) package and {{lump-sum}} (LS) package, which can be flexibly selected by prosumers based on their own preferences. Analytical energy-trading strategies will be derived for WP prosumers and LS prosumers based on non-cooperative games and Nash resource allocation strategies, respectively. In this work, a social cost optimization problem will be formulated for the EA, where the detailed WP/LS selection plans are unknown in advance. Consequently, a stochastic Stackelberg game between prosumers and the EA is formulated, and a two-level stochastic convex programming algorithm is proposed to minimize the expectation of the social cost. The performance of the proposed algorithm is demonstrated with a two-settlement based market model in the simulation.
\end{abstract}

\begin{IEEEkeywords}
Social cost optimization, energy prosumer, day-ahead market, energy balancing market, stochastic Stackelberg game, wholesale price, lump-sum price, game theory.
\end{IEEEkeywords}

\IEEEpeerreviewmaketitle
\section{Introduction}

\subsection{Background and Motivation}\label{s1}

\IEEEPARstart{M}{odern} wholesale electricity market usually contains a two-settlement process: {{day-ahead}} (DA) price settlement with DAM and {{real-time}} (RT) price settlement with EBM. In DAM, an independent {{system operator}} (ISO) collects bids and offers from market participants and clears the DAM. This task is especially important because of the difficulty in storing energy on a large scale and the high cost associated with supply failure \cite{hameer2015review}. In addition, to mitigate the imbalance between supply and demand in RT horizon, there is a final balancing process carried out between certain {{balance responsible party}} (BRP) (e.g., energy consumers) and the {{transmission system operator}} (TSO) in EBM.
This kind of market framework has been widely applied, such as in California electricity market and Nordic electricity market.
Therefore, in the two-settlement framework, it is beneficial for the BRPs to seek an optimal tradeoff between DAM and EBM to reduce their total consumption costs \cite{21}.

With such an objective, in this work, we consider a cluster of energy prosumers equipped with WPPs to support (part of) their energy demands. A common EA is established to provide the gateways for prosumers to trade with DAM and EBM. The EA aims to optimize the social energy consumption cost by making pricing strategies on the prosumers. Specifically, two price packages, namely WP package and LS package, are provided by the EA to meet the different preferences of prosumers. Wholesale pricing has been proved to be a mature pricing scheme that widely applied in the modern electricity market, e.g., studied in \cite{xu2015efficient,parvania2013optimal}. However, in many commodity markets, the discussion on lump-sum prices is still limited. In a lump-sum market, the commodity buyers would like to pay a total payment to the commodity providers in advance, and their demands will be satisfied in spite of some unforeseen risks. The advantages of LS prices conclude: (i) reducing the trading costs related to the commodity verification and measurement, (ii) simplifying the payment process, and (iii) motivating commodity providers to implement more economical strategies to fulfill the demand of commodity buyers, etc \cite{khalafalla2018unit}. In this work, we will explore a situation where the wholesale price and lump-sum price co-exist and can be flexibly selected by prosumers. In addition, we consider a dual-price principle for EBM, where the price of balancing energy will be affected by the imbalance directions. Under such a framework, we will investigate the following three questions.
\begin{itemize}
  \item How to model the selections of prosumers between WP and LS packages?
  \item What's the equilibrium (if exists) by considering the non-cooperative behaviours of prosumers?
  \item How to make an economical pricing strategy for the WP and LS packages to optimize the social cost of the community?
\end{itemize}

\subsection{Literature Review}
To increase the economic profit of BRPs and mitigate the imbalance of power systems, various optimization strategies have been investigated in the existing works \cite{vagropoulos2013optimal,iria2017trading,kohansal2016price,rahimiyan2016strategic,cui2018two,pei2016optimal}.
For example, in \cite{vagropoulos2013optimal}, an electric vehicle aggregator was established to minimize the social cost generated in both DAM and EBM by setting incentive prices. The optimization problem was formulated as a two-stage stochastic linear programming problem which takes the uncertainty of market conditions into account.
The authors of \cite{iria2017trading} discussed an optimal bidding problem for prosumers by considering the flexible loads in the power system. A stochastic programming problem was formulated to minimize the expected social cost generated in both DA and RT horizons. An optimal scheduling problem with time-shiftable loads was proposed in \cite{kohansal2016price}. To minimize the total consumption cost of the loads in both DAM and EBM, a mixed-integer programming algorithm with an accelerated computational speed was proposed. The authors of \cite{rahimiyan2016strategic} considered a virtual power plant which consists of WPPs and energy storage devices. A robust bidding algorithm was proposed for the virtual plant to maximize its profit by considering the uncertain output of WPPs. 

In addition, to model the non-cooperative behaviours of market participants, game theoretic methods have been widely studied in recent years. In game problems, each agent makes the best strategy for itself by observing the strategy of its rivals. Therefore, it's of great significance to establish certain coordinator (aggregator) to influence the behaviour of the agents such that the social benefit is optimized \cite{8}. The authors of \cite{bayes} argued that, in the electricity market, the price regulator can impose compulsory emission constraints on power generation entities to harvest long-term economic and environmental benefits by analysing the NE of the entities.
The authors of \cite{wang2015game} considered a game-based distributed loss reduction allocation problem for utility companies in the power system. An optimal marginal pricing strategy was proposed for the company community to maximize the overall profit. In \cite{esfahani2018multiagent}, a double auction game problem was proposed among loads, generators and energy storage devices. A market manager was established to optimize the social profit by making pricing strategies based on NE. In \cite{maharjan2013dependable}, the equilibrium among energy suppliers and consumers was analyzed by Nash and Stackelberg games. By modelling the impact of system attackers, it shows that the social profit and the reliability of the system can be improved by establishing adequate energy reserves. In \cite{wang2019noncooperative}, a non-cooperative social welfare optimization problem was studied for a batch of loads and a common load aggregator. By reformulating the social welfare optimization problem as a potential game, a distributed penalty based spatial adaptive play algorithm was proposed for all the agents, which induces the NE to the social optimal solution.


Different from the aforementioned works, the new features of this work are twofold. Firstly, we propose a future two-settlement based electricity market model with two optional price packages: WP package and LS package, which can be selected by the prosumers flexibly. Compared with single price package market, e.g., discussed in \cite{bayes,wang2015game,esfahani2018multiagent,maharjan2013dependable,wang2019noncooperative}, two price package can provide more choices for prosumers and adapt to their different preferences. Secondly, as pointed out in many research works, the dual-price principle of EBM may introduce discontinuous and/or non-convex characteristics into the market and result in non-deterministic or sub-optimal solutions \cite{o2016dual,ye2015transmission,ruiz2012pricing}. To address this issue, based on the stochastic package selections of the prosumers, it can be revealed that the mathematical prototype of the formulated social cost optimization problem is essentially a stochastic Stackelberg game with the EA being the leader and prosumers being the followers \cite{chang1983stochastic}. To solve this problem, we propose a two-level stochastic convex programming algorithm based on space-partitioning method.


The contributions of this paper are summarized as follows.
\begin{itemize}
  \item We propose a two-settlement based electricity market model which consists of a cluster of energy prosumers and a common EA. In this market, the entire energy-trading scheme of the prosumers is designed by a two-period process, namely {{package-bidding}} (PB) period and {{quantity-bidding}} (QB) period. In PB period, prosumers are allowed to make selections between WP and LS packages flexibly based on their own preferences, which can be modelled by a stochastic process. Then, in QB period, WP prosumers make detailed energy-purchasing strategies in DAM and EBM with wholesale prices, and the EA will make energy allocation plans for LS prosumers. By considering the Nash games among WP prosumers and the Nash resource allocation strategies for LS prosumers, the analytical optimal solutions for WP and LS prosumers are derived.
  \item Based on the NE analysis of WP and LS prosumers and the dual-price principle of EBM, we show that the resulting social cost function is non-deterministic with respect to the pricing strategy of the packages. To highlight the features of the proposed market model, we extract the mathematical prototype of the overall optimization architecture, which essentially follows the concept of stochastic Stackelberg game. Finally, a two-level stochastic convex programming is proposed to minimize the expected social cost. The ramp rate limits of EBM and the budget recovery of the EA are considered to ensure the reliability of the power system and the sustainability of the EA's service.
  \item Some principles and advantages of the proposed price-package selection scheme will be discussed. The performance of the proposed optimization algorithm is illustrated in the simulation by using the price data of Finland power grid.
\end{itemize}

The rest of the paper is organized as follows. {Section \ref{a}} presents the framework of the proposed market and the mathematical models of WPPs, prosumers and EA. In {Section \ref{sa}}, the optimization problems for WP and LS prosumers are formulated. Analytical optimal strategies of prosumers are derived based on non-cooperative games and Nash resource allocations. Then, the social cost function of the community is formulated based on the NE among the prosumers. In {Section \ref{c}}, a stochastic Stackelberg game is formulated between the prosumers and EA, and a two-level stochastic convex programming algorithm is proposed for the EA to minimize the expected social cost of the prosumer community. In Section \ref{cd}, the principles and advantages of the proposed price-package based optimization scheme are explained. The simulation is conducted in {Section \ref{d}}. {Section \ref{s5}} concludes this paper.

\section{System Modeling}\label{a}
\begin{figure}[htpb]
  \centering
  \includegraphics[height=5cm,width=8cm]{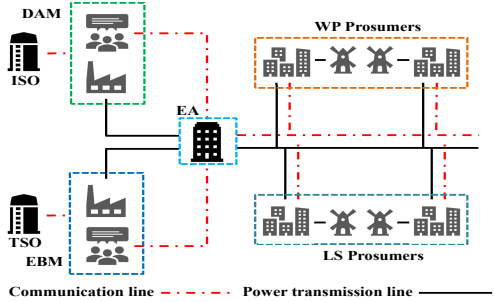}\\
   \caption{An illustrative framework of the proposed electricity market.}\label{microgrid}
\end{figure}

In this paper, we consider a two-settlement based electricity market consisting of a prosumer community $\mathcal{N}:=\{1,...,N\}$ and a common EA over time horizon $\mathcal{T}$. In this market, each prosumer is equipped with a local WPP to support (part of) its energy demand. The EA aims to optimize the social cost of the community by providing two price packages for the prosumers, namely WP package and LS package.
The framework of the proposed market is illustrated in Fig. \ref{microgrid}.

\subsection{Market Model}

In our problem, the energy-trading process is completed through a two-settlement process with DAM and EBM.

\subsubsection{DAM Model} The energy resource of DAM is assumed to be a thermal generator, and the generation cost is modeled by a quadratic function
\begin{equation}\label{}
G({d}_{tot,t}) := \frac{a}{2}{d}_{tot,t}^2+ bd_{tot,t}+c,
\end{equation}
where $a>0$, $b \geq 0$, $c\geq 0$, and ${d}_{tot,t} \geq 0$ is the total demand of prosumers in DAM. \footnote[1]{In this paper, time instant $t$ means the $t$th energy consumption slot in RT horizon.} Then, the price of DA energy can be determined by the marginal cost of $G({d}_{tot,t})$, which is
\begin{equation}\label{d1}
  P_t :=G'({d}_{tot,t}) = ad_{tot,t} + b.
\end{equation}

\subsubsection{EBM Model} According to the articles of Commission Regulation (EU), the price of balancing energy can be determined by a dual-price principle for different balancing regions and hours \cite{euets}.
Hence, in our model, the unit price of balancing energy is defined by
\begin{align}\label{sa1}
C_{B,t}:=
\left\{
  \begin{array}{ll}
    C_{UR,t}, & \hbox{if $x_{tot,t} \geq 0,$} \\
    C_{DnR,t}, & \hbox{if $x_{tot,t} < 0,$}
  \end{array}
\right.
\end{align}
where $x_{tot,t}$ is the total quantity of balancing energy. $x_{tot,t} \geq 0$ means the whole prosumer community requires a quantity $x_{tot,t}$ of balancing energy from EBM, and $x_{tot,t} < 0$ means the community injects a quantity $|x_{tot,t}|$ of surplus energy into EBM. $C_{UR,t}$ and $C_{DnR,t}$ are the up- and down-regulation prices, respectively, which are determined by the TSO. $C_{B,t}$ can be positive, negative or zero depending on the imbalance settlement periods, areas and imbalance directions \cite{euets}.

\subsection{Wind Power Plants}\label{rra}
Due to the uncertainty of wind, we employ the beta distribution to model the probability density function (PDF) of the random output of WPPs. The PDF of WPP $i$ (i.e., the WPP owned by the $i$th prosumer) can be written as \cite{bludszuweit2008statistical}
\begin{align}\label{ff}
    f_{i,t}(w_{i,t}):= \frac{1}{\mathcal{B}(\alpha_{i,t},\beta_{i,t})}(\frac{w_{i,t}}{w^{ca}_i})^{\alpha_{i,t}-1} (1-\frac{w_{i,t}}{w^{ca}_i})^{\beta_{i,t}-1},
\end{align}
where $\mathcal{B}(\alpha_{i,t},\beta_{i,t}) =\frac{\Gamma(\alpha_{i,t}) \Gamma(\beta_{i,t})}{\Gamma(\alpha_{i,t}+\beta_{i,t})}$ and $\Gamma(\cdot)$ is a Gamma function, $\alpha_{i,t},\beta_{i,t}> 1$, $i\in \mathcal{N}$. $w_{i,t}\geq 0$ is an independent and identically distributed random variable (output) of WPP $i$. $w^{ca}_i>0$ is the capacity of WPP $i$. In our problem, we assume that the PDF $f_{i,t}(w_{i,t})$ is open data and can be accessed by all the agents in the market. For convenience purpose, we define parameters $\mu_{i,t}:=\frac{\alpha_{i,t}}{\alpha_{i,t}+\beta_{i,t}}$ and $\nu^2_{i,t}:= \frac{\alpha_{i,t}\beta_{i,t}}{(\alpha_{i,t} + \beta_{i,t})^2 (\alpha_{i,t} + \beta_{i,t}+1)}$, which are the mean and variance of $w_{i,t}$, respectively \cite[pp. 424]{gupta2004handbook}. In the rest of this paper, operation $\int_0^{w^{ca}_i} (\cdot) \mathrm{d}w_{i,t}$ is written as
$\int_i (\cdot) \mathrm{d}w_{i,t}$ for simplicity.

\subsection{EA and Prosumers}\label{11}

In this market, the energy supply of prosumers comes from three parts: DAM, EBM and WPPs. Hence, the energy balance equation for prosumer $i$ can be expressed by
 \begin{equation}\label{ebe}
  u_{i,t}=x_{i,t}+d_{i,t}+w_{i,t}.
\end{equation}
In (\ref{ebe}), $u_{i,t}$ is the planned consumption quantity of prosumer $i$. $d_{i,t}$ and $x_{i,t}$ are energy-purchasing quantities of prosumer $i$ in DAM and EBM, respectively. Thus, the total demand of DA energy can be written as $d_{tot,t} = \sum_{i\in \mathcal{N}} d_{i,t}$, and the total demand of balancing energy is $x_{tot,t} = \sum_{i\in \mathcal{N}} x_{i,t}$. Since (\ref{ebe}) should hold at all times, in the following discussion, we set $x_{i,t}$ as the independent variable, and $d_{i,t}$ can be represented by
\begin{equation}\label{d2}
  d_{i,t} = u_{i,t} - w_{i,t} -x_{i,t}.
\end{equation}

For each prosumer, there are two package choices: WP and LS packages. In WP package, the payment of prosumers is decided by the energy-purchasing quantities and the cleared unit price. In LS package, each prosumer pays a lump-sum payment for the energy consumption. Finally, the overall energy demand will be sent to DAM and EBM under the management of EA. In the proposed mechanism, the pricing strategies of WP and LS packages for prosumer $i$ are designed as
\begin{equation}\label{cad}
\mathcal{Y}_{i,t}:=\left\{
            \begin{array}{ll}
                      B_{i,t}, &\hbox{if $i \in \mathcal{N}^{LS}$,} \\
                      R_t^{WP}x_{i,t}+P_td_{i,t},  & \hbox{if $i \in \mathcal{N}^{WP}$,}
                    \end{array}
                  \right.
\end{equation}
where $B_{i,t}$ is the LS price if prosuemr $i$ chooses LS package and $R_t^{WP}$ is a uniform unit price of balancing energy in WP package. $\mathcal{N}^{WP}$ and $\mathcal{N}^{LS}$ are the sets of prosumers who select WP package and LS package, respectively. Obviously, $\mathcal{N}^{WP} \cap \mathcal{N}^{LS} = \emptyset$ and $\mathcal{N}^{WP} \cup \mathcal{N}^{LS} = \mathcal{N}$. A positive $R_t^{WP}$ means a charge to WP prosumers when they need additional energy from EBM (i.e., $x_{i,t} \geq 0$, $i \in \mathcal{N}^{WP}$). If prosumers inject excessive energy into EBM (i.e., $x_{i,t} < 0$, $i \in \mathcal{N}^{WP}$), they can receive rewards from EA. $P_td_{i,t}$ is the energy cost of certain WP prosumer $i$ in DAM with demand $d_{i,t}$. In (\ref{cad}), $B_{i,t}$ and $R_t^{WP}$ are independent pricing strategies to be determined by the EA.

\subsection{Poisson Binomial Distribution Model of WP/LS Package Selections}
Due to the stochastic behaviours of prosumers in PB period, the WP/LS selections are modelled by Poisson binomial distribution, where the probability of choosing WP package of prosumer $i$ is defined by $q_{i} \in [0,1]$\footnote[2]{In this work, $q_{i}$ can be determined with historical data and assumed to be time-invariant in $\mathcal{T}$, $i \in \mathcal{N}$.}. In an $N$-prosumer community, there are $2^{N}$ selection scenarios in total. Hence, the probability of an $n$-WP prosumer scenario can be calculated by
\begin{equation}\label{poi}
  Q(n) :=\sum_{\mathcal{A}_{n,\mathcal{N}} \in {\mathcal{G}^{\mathcal{N}}_{n}}}\prod_{i \in \mathcal{A}_{n,\mathcal{N}}}q_{i} \prod_{j \in \mathcal{A}_{n,\mathcal{N}}^c} (1-q_j), \quad n \in \mathcal{N}^+,
\end{equation}
where $\mathcal{N}^+= \{0,1,2,...,N\}$.
${\mathcal{G}^{\mathcal{N}}_{n}}$ is a set where each element is composed of $n$ prosumers belonging to $\mathcal{N}$. $\mathcal{A}_{n,\mathcal{N}}$ is an element of ${\mathcal{G}^{\mathcal{N}}_{n}}$ and $\mathcal{A}^c_{n,\mathcal{N}}$ is the complement of $\mathcal{A}_{n,\mathcal{N}}$ in $\mathcal{N}$, i.e., $\mathcal{A}^c_{n,\mathcal{N}}= \mathcal{N} \setminus \mathcal{A}_{n,\mathcal{N}} $ \cite{wang1993number}.

%

\section{Cost Functions And Nash Equilibrium Analysis}\label{sa}

In this section, we will derive the equilibrium of prosumers under different packages and further formulate the social cost function of the prosumer community.




\subsection{Cost Function of WP Prosumers}\label{wp}

The energy prices of DAM and EBM are assumed to be accessible and predictable in DA horizon, respectively, by prosumers. Also, WP prosumers are assumed to be rational and focus on optimizing their own benefits ``selfishly''. Specifically, due to the uncertainty of the output of WPPs, WP prosumers play games to minimize their own expected consumption costs by determining the energy-purchasing strategies in DAM and EBM. The cost function of WP prosumer $i$ is defined as the expected total cost generated in both DAM and EBM, which is
\begin{align}\label{uh}
  {U}_{i,t} & (x_{i,t},  \bm{x}_{-i,t};  R_t^{WP})  :=   \mathbb{E}[R_t^{WP}x_{i,t} + P_t d_{i,t}] \nonumber \\
  = &  R_t^{WP}x_{i,t}+ \mathbb{E} [P_t d_{i,t}] \nonumber \\
  =  & R_t^{WP}x_{i,t} +  \mathbb{E}[(u_{i,t} - w_{i,t} - x_{i,t}) \nonumber \\
& \cdot (a \sum_{l \in \mathcal{N}}(u_{l,t} - w_{l,t} - x_{l,t}) + b )], \quad i \in \mathcal{N}^{WP},
\end{align}
where $\bm{x}_{-i,t}=(x_{1,t},...,x_{i-1,t},x_{i+1,t},...,x_{N,t})^T \in \mathbb{R}^{N-1}$ and $\mathbb{E}[\cdot]$ is the expectation operator. In (\ref{uh}), the third equality uses formulas (\ref{d1}) and (\ref{d2}).

\subsection{Expense Elaboration Equations for LS Prosumers}\label{vp}

In the following, we will derive a fair LS pricing strategy for EA by elaborating the components of LS price and explain how to realize the energy allocation from DAM and EBM to LS prosumers. For time $t$, an LS prosumer $i \in \mathcal{N}^{LS}$ pays an LS price, $B_{i,t}$, to EA and receives the energy product without caring about how the energy is delivered. In the perspective of the EA, allocating the energy from electricity markets to LS prosumers can be regarded as a resource allocation problem, where the costs generated in this process will influence the final LS prices. In a single-settlement market with static demand, the lump-sum price of certain participant can be proportional to its demand \cite{waslander2007convergence}. However, in a two-settlement market with elastic prices, it is reasonable to derive an equilibrium, where none of the participants would like to change its strategy unilaterally. Therefore, in this work, the allocation strategy for LS prosumers will be determined by Nash resource allocation approach \cite{bredin2000game1,johari2005game,xu2014game}, which can characterize the self-centric intention of the LS prosumers. Towards this end, we introduce a price component $R^{LS}_t$, which reflects the cost of balancing service associated with EBM.
Essentially, $R^{LS}_t$ is not a real charge to prosumers but a price component that decides the final LS price. This settlement coincides with the fact that the final electricity price can be influenced by some immeasurable factors, such as the choice of market participants \cite{szkuta1999electricity}. In our problem, the cost of balancing service for LS prosumer $i$ is defined as $S_{i,t}:=R^{LS}_t x_{i,t}$.
In a reasonable scenario, the total price of LS prosumers should equal the service cost of trading with both DAM and EBM, which can be expressed by the following {{expense elaboration equation}} (EEE):
\begin{align}\label{eb}
   & B_{i,t}  =  S_{i,t}  + \mathbb{E} [P_t d_{i,t}] \nonumber \\
   & =  R_t^{LS}x_{i,t} +  \mathbb{E}[(u_{i,t} - w_{i,t} - x_{i,t}) \nonumber \\
& \cdot (a \sum_{l \in \mathcal{N}}(u_{l,t} - w_{l,t} - x_{l,t}) + b )], \quad i \in \mathcal{N}^{LS}.
\end{align}
At the right-hand side of (\ref{eb}), the first term indicates the cost of balancing service; the second term denotes the expected cost of DA energy, whose derivation is similar to (\ref{uh}). (\ref{eb}) implies that even though price ${R}_t^{LS}$ is not announced to LS prosumers, it does reflect the inner relationship between final LS price and the balancing energy to be consumed.
To be reasonable, the feasible ranges of $R_t^{WP}$ and $R_t^{LS}$ are set as $\underline{R}_t^{WP}>0$ and $\underline{R}_t^{LS} > 0$, respectively.
\begin{Assumption}\label{a1}
Assume that
\begin{align}\label{sf}
0 \leq b \leq \min\{\underline{R}_t^{WP}, \underline{R}_t^{LS} |\forall t \in \mathcal{T} \}.
\end{align}
\end{Assumption}

\subsection{Nash Equilibrium of WP Prosumers and Energy Allocation Strategy for LS prosumers}\label{sc}

Based on the previous discussion, the objective of certain prosumer $i$ is minimizing ${U_{i,t}}(x_{i,t},\bm{x}_{-i,t}; R_t^{WP})$ by (\ref{uh}) or $B_{i,t}$ by (\ref{eb}) depending on which package is selected. Hence, by putting all the prosumers into a common ``competition pool'', a uniform objective function can be obtained as
\begin{align}\label{uy}
 & {{U_{i,t}}(x_{i,t},\bm{x}_{-i,t};R_{i,t})}
    :=  {R}_{i,t} x_{i,t}  + \mathbb{E} [ P_t d_{i,t}] \nonumber \\
= &  R_{i,t}x_{i,t} + \mathbb{E} [  (u_{i,t} - x_{i,t} - w_{i,t}) (a\sum_{l \in \mathcal{N}} (u_{l,t} - x_{l,t}  \nonumber \\
    & - w_{l,t}) + b ) ] \nonumber  \\
 = &   R_{i,t}x_{i,t} + \mathbb{E} [ (u_{i,t} - x_{i,t} - w_{i,t})(a\sum_{z \in \mathcal{N} \setminus \{i\}} (u_{z,t} \nonumber\\
&   - x_{z,t} - w_{z,t}) + a (u_{i,t}  - x_{i,t} - w_{i,t}) + b ) ] \nonumber  \\
  = & R_{i,t}x_{i,t} + (u_{i,t} - x_{i,t} - \mathbb{E}[w_{i,t}]) \nonumber\\
&\cdot (a\sum_{z \in \mathcal{N} \setminus \{i\}}  (u_{z,t}  - x_{z,t} - \mathbb{E}[w_{z,t}]) + b )  +a (u_{i,t}  - x_{i,t})^2 \nonumber\\
& + a \mathbb{E}[w^2_{i,t}] - 2 a \mathbb{E}[w_{i,t}](u_{i,t}  - x_{i,t}) \nonumber\\
= & R_{i,t}x_{i,t} + (u_{i,t} - x_{i,t} - \mu_{i,t}) (a\sum_{z \in \mathcal{N} \setminus \{i\}} (u_{z,t}  - x_{z,t} \nonumber\\
& - \mu_{z,t})  + b )  +a (u_{i,t}  - x_{i,t})^2  + a ((w^{ca}_i)^2  \nonumber\\
& -2 ( w^{ca}_iCF_{i,t}(w^{ca}_i)- CCF_{i,t}(w^{ca}_i) )) \nonumber\\
&- 2 a \mu_{i,t}(u_{i,t}  - x_{i,t}),
\end{align}
where
\begin{align}\label{15}
R_{i,t}=\bigg\{\begin{array}{ll}
            R_t^{WP},  & \hbox{if $i \in \mathcal{N}^{WP}$,}\\
            R^{LS}_t, &\hbox{if $i \in \mathcal{N}^{LS}$,} \\
                    \end{array}
\end{align}
with $CF_{i,t}(w_{i,t})=\int_i F_{i,t}(w_{i,t}) \mathrm{d}w_{i,t}$, $CCF_{i,t}(w_{i,t})=\int_i CF_{i,t}(w_{i,t}) \mathrm{d}w_{i,t}$, where $F_{i,t}(w_{i,t}) = \int_i f_{i,t}(w_{i,t}) \mathrm{d}w_{i,t}$ (i.e., the cumulative density function of $w_{i,t}$).
In the fifth equality, we use $\mu_{i,t}= \mathbb{E}[w_{i,t}]$, $F_{i,t}(w^{ca}_{i}) = 1$, $F_{i,t}(0)=CF_{i,t}(0)=CCF_{i,t}(0)=0$.

In this work, we consider that, at any time $t$, the expected total demand of DA energy of the community should be non-negative, i.e., $\sum_{i \in \mathcal{N}} d_{i,t} = \sum_{i \in \mathcal{N}}(u_{i,t}-\mu_{i,t}) - \sum_{i \in \mathcal{N}} x_{i,t} \geq 0$, which means prosumers can not sell energy to DAM due to some practical reasons, e.g., the large power disturbances caused by wind resources \cite{kyritsis2017electricity}. Then, the optimization problem of the prosumers can be formulated as
\begin{align}
   \textbf{(P1)}: \quad & \min\limits_{x_{i,t}} \quad   {{U_{i,t}}(x_{i,t},\bm{x}_{-i,t}; R_{i,t})}, \quad \forall i\in \mathcal{N},  \nonumber \\
    & s.t. \quad  \sum_{i \in \mathcal{N}}(u_{i,t}-\mu_{i,t}) - \sum_{i \in \mathcal{N}} x_{i,t} \geq 0, \label{co1}\\
    & \quad \quad (\ref{15}), \quad  \forall i \in \mathcal{N}. \label{co2}
\end{align}
Note that (\ref{co2}) is settled in PB period prior to solving Problem (P1). In addition, we do not set any lower bound for $x_{i,t}$ since a negative solution of certain prosumer can imply injecting energy to BEM or other prosumers in the community.

To derive the equilibrium of Problem (P1), we define Lagrangian function $L_{i,t}(x_{i,t},\bm{x}_{-i,t},\lambda_t; R_{i,t}):= {U_{i,t}}(x_{i,t},\bm{x}_{-i,t}; R_{i,t}) + \lambda_t  (-g_t( \bm{x}_{t}))$, with $g_t( \bm{x}_{t}) = \sum_{i \in \mathcal{N}}(u_{i,t}-\mu_{i,t}) - \sum_{i \in \mathcal{N}} x_{i,t}$, $\lambda_t \in \mathbb{R}$ is the Lagrangian multiplier. Then, we can have $\nabla_{x_{i,t}} L_{i,t}(x_{i,t},\bm{x}_{-i,t},\lambda_t; R_{i,t}) = \nabla_{x_{i,t}} {U_{i,t}}(x_{i,t},\bm{x}_{-i,t},\lambda_t; R_{i,t}) - \lambda_t \nabla_{x_{i,t}} g_t( \bm{x}_{t})$.
\begin{Lemma}
(Generalized Nash Equilibrium \cite[Thm. 8]{facchinei2010generalized}) If ${U_{i,t}}(x_{i,t},\bm{x}_{-i,t};R_{i,t})$ is convex at $x_{i,t}$ with (\ref{co1}) convex and closed, $\forall i \in \mathcal{N}$, then $(\bm{x}^*_t, \lambda_t^*)$, which solves the system
\begin{align}\label{}
& \left(
  \begin{array}{c}
    \nabla_{x_{1,t}} L_{1,t}(x_{1,t},\bm{x}_{-1,t},\lambda_t; R_{1,t}) \\
    \vdots \\
    \nabla_{x_{N,t}} L_{N,t}(x_{N,t},\bm{x}_{-N,t},\lambda_t; R_{N,t}) \\
  \end{array}
\right) = \bm{0}, \label{b1} \\
& 0 \leq \lambda_t  \quad \bot \quad g_t( \bm{x}_{t}) \geq 0, \label{b2}
\end{align}
is called as an NE of Problem (P1), where $\bm{x}^*_{t}=(x^*_{1,t},...,x^*_{N,t})^T \in \mathbb{R}^N$.
\end{Lemma}


\begin{Theorem}\label{l1}
Suppose that Assumption \ref{a1} holds. Given that all prosumers have decided their price packages in PB period, the NE solved by {Problem (P1)} exists and is unique. Moreover, the analytical NE can be given by
\begin{align}
  & x^{WP*}_{h,t} =  u_{h,t} - \mu_{h,t} + \frac{b + \sum_{f \in \mathcal{N}, f \neq h}R_{f,t} -N{R}^{WP}_t}{a(N+1)}, \label{alr1} \\
  &  x^{LS*}_{j,t} =  u_{j,t} - \mu_{j,t} + \frac{b + \sum_{g \in \mathcal{N},g \neq j}R_{g,t} -N{R}_t^{LS}}{a(N+1)} , \label{alr2}
\end{align}
where $x^{WP*}_{h,t}$ and $ x^{LS*}_{j,t}$ are NE solutions for WP and LS prosumers, respectively, $h \in \mathcal{N}^{WP}$, $j \in \mathcal{N}^{LS}$, and parameter $R_{i,t}$ in (\ref{alr1}) and (\ref{alr2}) is determined by (\ref{15}), $i\in \mathcal{N}$. In addition, in an $n$-WP prosumer community, the total demand of balancing energy can be given by
\begin{align}\label{a2}
  & x^*_{n,tot,t} : = \frac{Nb - (n R_t^{WP} + (N-n) R_t^{LS})}{a(N+1)} + \sum_{i \in \mathcal{N}} (u_{i,t}-\mu_{i,t}).
\end{align}
\end{Theorem}

The proof can be referred to in Appendix \ref{l1p}. To achieve the NE, many existing decentralized algorithms can be applied, e.g., proposed in \cite{yi2017distributed,sun2018distributed}, which is not to be discussed in this work.



\begin{Remark}
In the proposed PB-QB mechanism, the following features worth emphasizing.
\begin{enumerate}
  \item The prosumers are free to choose any package according to their preferences, which are unknown by the EA in advance. This complies with the common sense in many commodity markets.
  \item For LS package, both the LS prosumers and EA will take the risk of the uncertainty of the output of WPPs, because the lump-sum contracts are usually signed before the execution period. In this sense, the package selection can also be a result of the evaluation of prosumers on the output of WPPs.
  \item Additional to the community-EBM energy-trading process managed by the EA, prosumers with surplus or inadequate energy are also allowed to trade energy within the community, which realizes a peer-to-peer energy-trading fashion \cite{anoh2019energy}.
  \item By integrating all the prosumers into a common market, the global NE for all the WP and LS prosumers can be obtained as shown in Theorem \ref{l1}. This reveals the fact that all the prosumers in the same market, regardless of which packages they choose, are potential competitors.
\end{enumerate}
\end{Remark}

\subsection{Social Cost Function of Prosumer Community}

In our problem, the demand of each prosumer is assumed to be known by EA in advance, which is a commonly used assumption in the works on aggregation strategies \cite{ishizaki2020day}.
Then, the objective function of EA can be defined by the expected social cost generated in both DAM and EBM, which is
\begin{align}\label{27}
 & W_t (\bm{x}_t; C_{B,t}) :=  \mathbb{E} (C_{B,t}x_{tot,t}+  P_td_{tot,t})\nonumber  \\
=  & C_{B,t} x_{tot,t}+ \mathbb{E} [  \sum_{i \in \mathcal{N}}(u_{i,t} - x_{i,t} - w_{i,t}) \nonumber \\
& \cdot (a\sum_{i \in \mathcal{N}} (u_{i,t} - x_{i,t} - w_{i,t}) + b ) ] \nonumber \\
= &  C_{B,t} x_{tot,t}  + \mathbb{E} [ a (\sum_{i \in \mathcal{N}}(u_{i,t}-w_{i,t}))^2 + a(\sum_{i\in \mathcal{N}} x_{i,t})^2 \nonumber \\
& - (2a \sum_{i\in \mathcal{N}} (u_{i,t} - w_{i,t}) + b ) \sum_{i\in \mathcal{N}} x_{i,t}  + b \sum_{i \in \mathcal{N}}(u_{i,t}-w_{i,t}) ] \nonumber \\
= & \mathbb{E} [ ax_{tot,t}^2 + (  C_{B,t} - 2a \sum_{i\in \mathcal{N}} (u_{i,t} - w_{i,t}) - b  ) x_{tot,t} \nonumber \\
&  + a (\sum_{i \in \mathcal{N}}u_{i,t} )^2 - 2a \sum_{i \in \mathcal{N}}u_{i,t} \sum_{i \in \mathcal{N}}w_{i,t} + a \sum_{i \in \mathcal{N}}w^2_{i,t} \nonumber \\
& +  a \sum_{i \in \mathcal{N}} \sum_{z \in \mathcal{N}\setminus \{i\}} w_{i,t} w_{z,t}   +b \sum_{i \in \mathcal{N}}(u_{i,t}-w_{i,t})] \nonumber \\
= & ax_{tot,t}^2 + (  C_{B,t} - 2a \sum_{i\in \mathcal{N}} (u_{i,t} - \mathbb{E} [w_{i,t}])  - b  ) x_{tot,t} \nonumber \\
&  + a (\sum_{i \in \mathcal{N}}u_{i,t} )^2 - 2a \sum_{i \in \mathcal{N}}u_{i,t} \sum_{i \in \mathcal{N}} \mathbb{E}[w_{i,t}] + a \sum_{i \in \mathcal{N}} \mathbb{E}[w^2_{i,t}] \nonumber \\
& +  a \sum_{i \in \mathcal{N}} \sum_{z \in \mathcal{N}\setminus \{i\}} \mathbb{E}[w_{i,t}] \mathbb{E}[w_{z,t}]   +b \sum_{i \in \mathcal{N}}(u_{i,t}- \mathbb{E}[w_{i,t}])  \nonumber \\
= & ax_{tot,t}^2 + (  C_{B,t} - 2a \sum_{i\in \mathcal{N}} (u_{i,t} - \mu_{i,t}) - b ) x_{tot,t} \nonumber \\
& + a (\sum_{i \in \mathcal{N}}u_{i,t} )^2 - 2a \sum_{i \in \mathcal{N}}u_{i,t} \sum_{i\in \mathcal{N}} \mu_{i,t}  + a \sum_{i \in \mathcal{N}} ( (w^{ca}_i)^2 \nonumber \\
& -2 ( w^{ca}_i CF_{i,t}(w^{ca}_i)- CCF_{i,t}(w^{ca}_i) ) ) \nonumber \\
& +  a  \sum_{i \in \mathcal{N}} \sum_{z \in \mathcal{N} \setminus \{i\}} \mu_{i,t} \mu_{z,t} +b \sum_{i \in \mathcal{N}}(u_{i,t}-\mu_{i,t})  \nonumber \\
 := & a x_{tot,t}^2 + \Phi_t x_{tot,t} + \Psi_t,
\end{align}
where $\bm{x}_t = (x_{1,t},...,x_{N,t})^T \in \mathbb{R}^N$,
\begin{align}\label{}
& \Phi_t =  C_{B,t}-2a\sum_{i \in \mathcal{N}}(u_{i,t}-\mu_{i,t})-b, \label{m1} \\
&  \Psi_t =  a (\sum_{i \in \mathcal{N}}u_{i,t} )^2 - 2a \sum_{i \in \mathcal{N}}u_{i,t} \sum_{i\in \mathcal{N}}\mu_{i,t}  \nonumber  \\
  & + a \sum_{i \in \mathcal{N}} ((w^{ca}_i)^2 -2 ( w^{ca}_i CF_{i,t}(w^{ca}_i)- CCF_{i,t}(w^{ca}_i) ) ) \nonumber  \\
  & +  a  \sum_{i \in \mathcal{N}} \sum_{z \in \mathcal{N}\setminus \{i\} } \mu_{i,t} \mu_{z,t} +b \sum_{i \in \mathcal{N}}(u_{i,t}-\mu_{i,t}).
\end{align}
In (\ref{m1}), $C_{B,t}$ is non-deterministic since the sign of $x_{tot,t}$ is non-deterministic (see (\ref{sa1})), which means (\ref{27}) can not be addressed by conventional deterministic optimization approaches. Therefore, we will discuss how to minimize the expectation of the social cost in the next section.

\section{Social Cost Optimization Strategy Development}\label{c}


\subsection{The Mathematical Prototype of the Problem}\label{v}

We will present our proposed social cost optimization strategy based on the following assumption.

\begin{Assumption}\label{as1}
{{(i)}} WP prosumers decide their energy-purchasing strategies by NE solution (\ref{alr1}).
{{(ii)}} The demands of LS prosumers are satisfied by Nash resource allocation principle (\ref{alr2}).
\end{Assumption}

Based on Assumption \ref{as1}, in an $n$-WP prosumer community, the total balancing energy quantity $x_{tot,t}$ in (\ref{27}) can be replaced by $x^*_{n,tot,t}$, which is a function of $(R^{WP}_t, R^{LS}_t)$ and $n$ as shown in (\ref{a2}). Therefore, the social cost function of an $n$-WP prosumer community can be reformulated as
\begin{align}\label{31}
& {W}_{n,t} ( R_t^{WP},  R_t^{LS} ; C_{B,t}) := a (x^*_{n,tot,t})^2 + \Phi_t x^*_{n,tot,t} + \Psi_t.
\end{align}
In addition, due to the stochastic $n$, the target of EA is designed to minimize the expectation of ${W}_{n,t}( R_t^{WP}, R_t^{LS} ; C_{B,t})$, which is
\begin{align}\label{32}
\mathbb{E} & [{W}_{n,t}  ( R_t^{WP} , R_t^{LS} ; C_{B,t})] \nonumber \\
& = \sum_{n \in \mathcal{N}^+}Q(n) {W}_{n,t}( R_t^{WP}, R_t^{LS} ; C_{B,t}).
\end{align}
Under such a framework, the interactions between EA and prosumers can be described by a two-stage incentive-based stochastic Stackelberg game discussed in \cite{chang1983stochastic}. The basic elements involved in this game are summarized in Table \ref{tm2}.
\begin{table}
\caption{Elements of the Stochastic Stakelberg Game between Prosumers and EA}\label{tm2}
\label{tab2}
\begin{center}
\begin{tabular}{ccc}
\hline
\multicolumn{3}{l}{Prosumers}  \\
\hline
BP & \tabincell{c}{Package Selection} & WP/LS Packages \\
\hline
& \tabincell{c}{Incentive \\ Received} & $R_{i,t} \in \{R_t^{WP}, R_t^{LS}\}$, $\forall i, \forall t$  \\
\cline{2-3}
 QP & \tabincell{c}{Optimization \\Problem} & $\min\limits_{x_{i,t}} {{U_{i,t}}(x_{i,t},\bm{x}_{-i,t} ;R_{i,t})}$, $\forall i, \forall t$  \\
\cline{2-3}
  & NE & $ x^{*}_{i,t}$, $\forall i, \forall t$ \\
\hline
\multicolumn{3}{l}{EA}  \\
\hline
  & \tabincell{c}{Optimization \\ Problem} & \tabincell{c}{$\min\limits_{R_t^{WP}, R_t^{LS}} \mathbb{E} [{W}_{n,t}( R_t^{WP}, R_t^{LS} ; C_{B,t})]$, $\forall t$}  \\
\cline{2-3}
  & Strategy & $(R_t^{WP}, R_t^{LS})$, $\forall t$  \\
\hline
\end{tabular}
\end{center}
\end{table}


\subsection{Determination of $C_{B,t}$}\label{sb}

Under Assumption \ref{as1}, $C_{B,t}$ can be determined by principle
\begin{align}\label{}
C_{B,t}=
\left\{
  \begin{array}{ll}
    C_{UR,t}, & \hbox{if $x^*_{n,tot,t} \geq 0,$} \\
    C_{DnR,t}, & \hbox{if $x^*_{n,tot,t} < 0.$}
  \end{array}
\right.
\end{align}
Then, with (\ref{a2}), it can be verified that if
\begin{align}\label{c1}
  n R_t^{WP} + (N-n) R_t^{LS} \leq  Nb
  +a (N+1) \sum_{i \in \mathcal{N}}(u_{i,t} - \mu_{i,t}),
\end{align}
then $x^*_{n,tot,t} \geq 0$ and $C_{B,t}=C_{UR,t}$; and if
 \begin{align}\label{c2}
  n R_t^{WP} + (N-n) R_t^{LS} >  Nb
  +a (N+1) \sum_{i \in \mathcal{N}}(u_{i,t} - \mu_{i,t}),
 \end{align}
then $x^*_{n,tot,t} < 0$ and $C_{B,t}=C_{DnR,t}$, $n \in \mathcal{N}^+$.
Hence, the value of $C_{B,t}$ is decided by number $n$ and $(R_t^{WP}, R_t^{LS})$. In the following, we will provide a space-partitioning method to determine the value of $C_{B,t}$ by discussing the location of $(R_t^{WP}, R_t^{LS})$ and $n$.

To explain how to determine $C_{B,t}$ when $n$ varies from $0$ to $N$, we introduce an auxiliary parameter $n_{\sigma} \in \mathcal{N}^+\setminus \{N\}$, where the balancing direction (i.e., the sign of $x^*_{n,tot,t}$) will be inverted when $n$ increases from $n_{\sigma}$ to $n_{\sigma}+1$ (including the trivial cases $x^*_{n_{\sigma},tot,t}=0$ and $x^*_{n_{\sigma+1},tot,t}=0$). Note that given any $(R_t^{WP}, R_t^{LS})$, (\ref{a2}) is an affine function of $n$. Therefore, with any pair $(R_t^{WP}, R_t^{LS})$, the value of $C_{B,t}$ with $n$ varying from $0$ to $N$ will be affected by the location of $n_{\sigma}$. In the following, we will discuss the value of $C_{B,t}$ under conditions $R_t^{LS} \geq R_t^{WP}$ and $R_t^{LS} < R_t^{WP}$ separately, and $C_{B,t}$ is written as $C^{\geq}_{B,t}(n_{\sigma},n)$ and $C^{<}_{B,t}(n_{\sigma},n)$ accordingly for clearness.
\subsubsection{Case 1}
If $R_t^{LS} \geq R_t^{WP}$, by (\ref{a2}), we have
\begin{align}\label{pp}
x^*_{n+1,tot,t} \geq x^*_{n,tot,t}, \quad \forall n \in \mathcal{N}^+ \setminus \{N\}.
\end{align}
Then, for certain inversion point $n_{\sigma}$ such that
\begin{align}\label{}
\underbrace{x^*_{N,tot,t}\geq ...\geq x^*_{n_{\sigma}+1,tot,t} \geq}_{\Gamma_1} 0\underbrace{ \geq x^*_{n_{\sigma},tot,t} \geq ... \geq x^*_{0,tot,t}}_{\Gamma_2}
\end{align}
(the situation $x^*_{N,tot,t} < ... < x^*_{n_{\sigma}+1,tot,t} < 0 < x^*_{n_{\sigma},tot,t}< ... < x^*_{0,tot,t}$ does not exist under condition $R_t^{LS} \geq R_t^{WP}$ since (\ref{pp}) should hold at all times), the range of $(R_t^{WP}, R_t^{LS})$ can be solved by
\begin{align}\label{}
 (R_t^{WP}, R_t^{LS}) \in {\mathcal{H}_{n_{\sigma},t}} : = & \mathcal{H}^{>}_{0,t} \cap ... \cap \mathcal{H}^{>}_{n_{\sigma},t} \nonumber \\
& \cap \mathcal{H}^{\leq}_{n_{\sigma+1},t} \cap ... \cap \mathcal{H}^{\leq}_{N,t},
\end{align}
where $\mathcal{H}^{>}_{0,t}$, $\mathcal{H}^{>}_{n_{\sigma},t}$, $\mathcal{H}^{\leq}_{n_{\sigma}+1,t}$ and $\mathcal{H}^{\leq}_{N,t} $ are the spaces defined by
\begin{align}\label{}
& \{(R_t^{WP}, R_t^{LS}) | 0*R_t^{WP} + N R_t^{LS} \nonumber \\
& \quad \quad \quad \quad \quad \quad  > Nb+a (N+1) \sum_{i \in \mathcal{N}}(u_{i,t} - \mu_{i,t})\}, \\
&  \{(R_t^{WP}, R_t^{LS}) |n_{\sigma} R_t^{WP} + (N-n_{\sigma}) R_t^{LS} \nonumber \\
& \quad \quad \quad \quad \quad \quad > Nb+a (N+1) \sum_{i \in \mathcal{N}}(u_{i,t} - \mu_{i,t})\},  \\
& \{(R_t^{WP}, R_t^{LS}) |(n_{\sigma}+1) R_t^{WP} + (N-n_{\sigma}-1) R_t^{LS} \nonumber \\
& \quad \quad \quad \quad \quad \quad \leq Nb +a (N+1) \sum_{i \in \mathcal{N}}(u_{i,t} - \mu_{i,t})\}, \\
& \{(R_t^{WP}, R_t^{LS}) |N R_t^{WP} + 0*R_t^{LS} \nonumber \\
& \quad \quad \quad \quad \quad \quad \leq Nb+a (N+1) \sum_{i \in \mathcal{N}}(u_{i,t} - \mu_{i,t})\},
\end{align}
respectively. Notation $(\cdot)^{\leq}$ means that the space is generated by letting the left-hand side of the inequality (\ref{c1}) less than or equal to the right-hand side with certain given value $n$. $(\cdot)^{>}$ means ``larger than'' accordingly. Essentially, the spaces $\mathcal{H}^{\leq}_{n_{\sigma}+1,t},..., \mathcal{H}^{\leq}_{N,t}$ are obtained by solving (\ref{c1}) (i.e., the condition of $x^*_{n,tot,t} \geq 0$ and $C^{\geq}_{B,t}(n_{\sigma},n)=C_{UR,t}$) by letting $n=n_{\sigma+1},...,N$, which means $\Gamma_1$ is satisfied; and $\mathcal{H}^{>}_{0,t},...,\mathcal{H}^{>}_{n_{\sigma},t} $ are obtained by solving (\ref{c2}) (i.e., the condition of $x^*_{n,tot,t} < 0$ and $C^{\geq}_{B,t}(n_{\sigma},n)=C_{DnR,t}$) by letting $n=0,...,n_{\sigma}$, which means $\Gamma_2$ is satisfied. \footnote[3]{Compared with $\Gamma_2$, constraint $x^*_{n,tot,t} < 0$ can be equivalent to $x^*_{n,tot,t} \leq 0$ in the sense that the solution constrained by the former can be sufficiently close to the latter if the constraint is activated.} In this case, given any $n_{\sigma}$ and space ${\mathcal{H}_{n_{\sigma},t}}$, the value of $C^{\geq}_{B,t}(n_{\sigma},n)$ is deterministic with $n=0,...,N$, i.e., $C^{\geq}_{B,t}(n_{\sigma},n)=C_{UR,t}$, $\forall$ $n \geq n_{\sigma}+1$, and $C^{\geq}_{B,t}(n_{\sigma},n)=C_{DnR,t}$, $\forall$ $n \leq n_{\sigma}$, $n \in \mathcal{N}^+$.

Note that by $R_t^{LS} \geq R_t^{WP}$, we can have $\mathcal{H}^{\leq}_{n_{\sigma+1},t} = \mathcal{H}^{\leq}_{n_{\sigma+1},t} \cap  ... \cap \mathcal{H}^{\leq}_{N,t}$ and  $\mathcal{H}^{>}_{n_{\sigma},t} = \mathcal{H}^{>}_{0,t} \cap ... \cap \mathcal{H}^{>}_{n_{\sigma},t}$.
Therefore, with any given $n_{\sigma}$, the value of $C^{\geq}_{B,t}(n_{\sigma},n)$ can be determined by the space with a simpler expression
\begin{equation}\label{}
  {{\mathcal{F}}_{{n}_{\sigma},t}}:= \mathcal{H}^{>}_{n_{\sigma},t} \cap \mathcal{H}^{\leq}_{n_{\sigma}+1,t},
\end{equation}
$\forall n \in \mathcal{N}^+$, $n_{\sigma} \in \mathcal{N}^+\setminus \{N\}$.

Furthermore, we define
\begin{align}\label{}
& {{\mathcal{F}}_{-1,t}}:=\{(R_t^{WP}, R_t^{LS}) | N R_t^{LS} \nonumber \\
& \quad \quad \quad \quad \quad \quad \leq Nb+a (N+1)\sum_{i \in \mathcal{N}}(u_{i,t} - \mu_{i,t})\}, \\
& {{\mathcal{F}}_{N,t}}:=\{(R_t^{WP}, R_t^{LS}) | N R_t^{WP} \nonumber \\
& \quad \quad \quad \quad \quad \quad> Nb+a (N+1)\sum_{i \in \mathcal{N}}(u_{i,t} - \mu_{i,t})\},
\end{align}
and define $C^{\geq}_{B,t}(-1,n)$ and $C^{\geq}_{B,t}(N,n)$ as the values of $C_{B,t}$ in ${{\mathcal{F}}_{-1,t}}$ and ${{\mathcal{F}}_{N,t}}$, respectively. Under the condition $R_t^{LS} \geq R_t^{WP}$, spaces ${{\mathcal{F}}_{-1,t}}$ and ${{\mathcal{F}}_{N,t}}$ ensure that $C^{\geq}_{B,t}(-1,n) = C_{UR,t} $ (i.e., $x^*_{n,tot,t} \geq 0$) and $C^{\geq}_{B,t}(N,n) = C_{DnR,t}$ (i.e., $x^*_{n,tot,t} < 0$), respectively, $\forall n \in \mathcal{N}^+$.

With the above arrangement, the whole space $\{(R_t^{WP}, R_t^{LS}) | R_t^{LS} \geq R_t^{WP} \}$ is split into $N+2$ sub-spaces, i.e., $\cup_{{n}_{\sigma} \in \mathcal{N}^{\pm}}{{\mathcal{F}}_{{n}_{\sigma},t}} = \{(R_t^{WP}, R_t^{LS}) | R_t^{LS} \geq R_t^{WP} \}$ and $\cap_{{n}_{\sigma} \in \mathcal{N}^{\pm}}{{\mathcal{F}}_{{n}_{\sigma},t}} = \emptyset$, where $\mathcal{N}^{\pm}=\{-1,0,...,N\}$ with $|\mathcal{N}^{\pm}| = N+2$. The value of $C^{\geq}_{B,t}(n_{\sigma},n)$ is deterministic within sub-space ${{\mathcal{F}}_{{n}_{\sigma},t}}$, $\forall n \in \mathcal{N}^+$, $n_{\sigma} \in \mathcal{N}^{\pm}$.
\subsubsection{Case 2}
By following the same logic, the value of $C^{<}_{B,t}(n_{\sigma},n)$ can be also obtained by splitting $\{(R_t^{WP}, R_t^{LS}) | R_t^{LS} < R_t^{WP} \}$ into $N+2$ sub-spaces, which are defined by
\begin{align}\label{}
& \bar{{\mathcal{F}}}_{-1,t}:= \{(R_t^{WP}, R_t^{LS}) | N R_t^{LS} \nonumber \\
& \quad \quad \quad \quad> Nb+a (N+1) \sum_{i \in \mathcal{N}}(u_{i,t} - \mu_{i,t})\},  \\
& \bar{{\mathcal{F}}}_{N,t}:= \{(R_t^{WP}, R_t^{LS}) | N R_t^{WP} \nonumber \\
& \quad \quad \quad \quad \leq Nb+a (N+1) \sum_{i \in \mathcal{N}}(u_{i,t} - \mu_{i,t})\},  \\
& \bar{{\mathcal{F}}}_{n_{\sigma},t}: = \mathcal{H}^{\leq}_{n_{\sigma},t} \cap \mathcal{H}^{>}_{n_{\sigma}+1,t}, \quad n_{\sigma} \in \mathcal{N}^+ \setminus \{N\},
\end{align}
where $\mathcal{H}^{\leq}_{n_{\sigma},t}$ and $\mathcal{H}^{>}_{n_{\sigma}+1,t}$ are the spaces determined by \begin{align}\label{}
&  \{(R_t^{WP}, R_t^{LS}) |n_{\sigma} R_t^{WP} + (N-n_{\sigma}) R_t^{LS} \nonumber \\
& \quad \quad \quad \quad \quad \quad \leq Nb+a (N+1) \sum_{i \in \mathcal{N}}(u_{i,t} - \mu_{i,t})\}, \\
& \{(R_t^{WP}, R_t^{LS}) |(n_{\sigma}+1) R_t^{WP} + (N-n_{\sigma}-1) R_t^{LS} \nonumber \\
& \quad \quad \quad \quad \quad \quad > Nb +a (N+1) \sum_{i \in \mathcal{N}}(u_{i,t} - \mu_{i,t})\},
\end{align}
respectively. Similar to Case 1, define $C^{<}_{B,t}(-1,n)$ and $C^{<}_{B,t}(N,n)$ as the values of $C_{B,t}$ in ${\bar{\mathcal{F}}_{-1,t}}$ and ${\bar{\mathcal{F}}_{N,t}}$, respectively. In this case, we can have that $C^{<}_{B,t}(-1,n) = C_{DnR,t}$ if $(R_t^{WP}, R_t^{LS}) \in \bar{\mathcal{F}}_{-1,t}$, $\forall n \in \mathcal{N}^+$; $C^{<}_{B,t} (n_{\sigma},n)= C_{DnR,t}$ if $(R_t^{WP}, R_t^{LS}) \in  \bar{\mathcal{F}}_{n_{\sigma},t}$ with $n_{\sigma}+1 \leq n \leq N$; $C^{<}_{B,t}(N,n) = C_{UR,t}$ if $(R_t^{WP}, R_t^{LS}) \in \bar{\mathcal{F}}_{N,t}$, $\forall n \in \mathcal{N}^+$; $C^{<}_{B,t} (n_{\sigma},n) = C_{UR,t}$ if $(R_t^{WP}, R_t^{LS}) \in \bar{\mathcal{F}}_{n_{\sigma},t}$ with $0 \leq n \leq n_{\sigma}$, $n_{\sigma} \in \mathcal{N}^+ \setminus \{N\}$. The discussion procedure is similar to Case 1 and is omitted for simplicity.

\begin{table}
\caption{Determination of $C_{B,t}(n_{\sigma},n)$ in terms of sub-spaces and $n$}\label{t1}
\begin{center}
\begin{tabular}{cccc}
  \hline
   & Sub-space & Range of $n$ & $C_{B,t}$ \\
   \hline
  \multirow{4}{*}{$R_t^{LS} \geq R_t^{WP}$} & $\mathcal{F}_{-1,t}$ & $[0,N]$ & $C_{UR,t}$ \\
  \cline{2-4}
   & \multirow{2}{*}{$\mathcal{F}_{n_{\sigma},t}$}  & $[0 ,{n}_{\sigma}]$ & $C_{DnR,t}$ \\
   \cline{3-4}
   & & $[{n}_{\sigma}+1,  N]$ & $C_{UR,t}$ \\
   \cline{2-4}
   & $\mathcal{F}_{N,t}$ & $[0,N]$ & $C_{DnR,t}$ \\
  \hline
  \multirow{4}{*}{$R_t^{LS} < R_t^{WP}$} & $\bar{\mathcal{F}}_{-1,t}$ & $[0,N]$ & $C_{DnR,t}$ \\
  \cline{2-4}
   & \multirow{2}{*}{$\bar{\mathcal{F}}_{n_{\sigma},t} $}  & $[0, {n}_{\sigma}]$ & $C_{UR,t}$ \\
   \cline{3-4}
   &  & $[{n}_{\sigma}+1, N]$ & $C_{DnR,t}$ \\
   \cline{2-4}
   & $\bar{\mathcal{F}}_{N,t}$ & $[0,N]$ & $C_{UR,t}$ \\
  \hline
      \multicolumn{4}{l}{$n_{\sigma} \in \mathcal{N}^+\setminus \{N\}$, $n\in \mathcal{N}^+$, $t \in \mathcal{T}$.} \\
       \hline
\end{tabular}
\end{center}
\end{table}


By Cases 1 and 2, the full 2-dimensional Euclidean space of $(R^{WP}_t, R^{LS}_t)$ is split into $2|\mathcal{N}^{\pm}| = 2N+4$ sub-spaces with
\begin{align}\label{}
& \cup_{{n}_{\sigma} \in \mathcal{N}^{\pm}}{{\mathcal{F}}_{{n}_{\sigma},t}} \cup_{{n}_{\sigma} \in \mathcal{N}^{\pm}}  \bar{{\mathcal{F}}}_{{n}_{\sigma},t} = \mathbb{R}^2, \\
&  \cap_{{n}_{\sigma} \in \mathcal{N}^{\pm}}{{\mathcal{F}}_{{n}_{\sigma},t}} \cap_{{n}_{\sigma} \in \mathcal{N}^{\pm}}  \bar{{\mathcal{F}}}_{{n}_{\sigma},t} = \emptyset.
\end{align}
The overall principle of the determination of $C_{B,t}$ is listed in {Table \ref{t1}}, by which the expression of ${W}_{n,t}(R_t^{WP}, R_t^{LS} ; C_{B,t})$ is deterministic in each sub-space, $\forall n \in \mathcal{N}^+$, $t \in \mathcal{T}$.

\subsection{Budget Recovery of EA}

In this subsection, we discuss the condition which recovers the expectation of the budget of EA. The net profit of EA is decided by the income and payment generated in the trading with prosumers, DAM and EBM. For any $n$-WP prosumer set $\mathcal{N}^{WP}_n \subseteq \{ \mathcal{N}^{WP} \subseteq \mathcal{N} | |\mathcal{N}^{WP}| = n \}$ and the corresponding LS prosumer set $\mathcal{N}^{LS}_n := \mathcal{N} \setminus \mathcal{N}^{WP}_n$, by (\ref{alr1}) and (\ref{alr2}), the total balancing energy of WP and LS prosumers satisfies
\begin{align}
 & \sum_{h \in \mathcal{N}^{WP}_n} x^{WP*}_{h,t} =  \frac{nb + n(N-n)(R_t^{LS} - R_t^{WP})-n R_t^{WP}}{a(N+1)} \nonumber \\
& \quad \quad  + \sum_{h \in \mathcal{N}^{WP}_n}(u_{h,t} - \mu_{h,t}) \nonumber \\
 & \geq  \frac{nb + n(N-n)(R_t^{LS} - R_t^{WP}) -n R_t^{WP}}{a(N+1)},  \nonumber \\
& \quad \quad  + n \min_{i \in \mathcal{N}}\{u_{i,t} - \mu_{i,t}\}
 : =  x^{WP*}_{n,tot,t}, \label{e1} \\
& \sum_{j \in \mathcal{N}^{LS}_n}  x^{LS*}_{j,t} \nonumber \\
 & =  \frac{(N-n)b + n(N-n)(R_t^{WP} - R_t^{LS})-(N-n)R_t^{LS}}{a(N+1)} \nonumber \\
 & \quad \quad + \sum_{j \in \mathcal{N}^{LS}_n}(u_{j,t} - \mu_{j,t}) \nonumber \\
 & \geq \frac{(N-n)b + n(N-n)(R_t^{WP} - R_t^{LS})-(N-n)R_t^{LS}}{a(N+1)} \nonumber \\ & \quad \quad +(N-n)\min_{i\in \mathcal{N}}\{u_{i,t} - \mu_{i,t}\}:=x^{LS*}_{n,tot,t}, \label{e2}
\end{align}
where (\ref{alr1}) and (\ref{alr2}) are used. Then, the net profit of the EA can be calculated by
\begin{align}\label{l}
 Z_{n,t} & (R^{WP}_t,R^{LS}_t,n_{\sigma})
: = ( \sum_{j \in \mathcal{N}^{LS}_n} B_{j,t} + R^{WP}_t \sum_{h \in \mathcal{N}^{WP}_n} x^{WP*}_{h,t} \nonumber  \\
&  +  \sum_{h \in \mathcal{N}^{WP}_n} \mathbb{E}[ (u_{h,t} - w_{h,t} -  x^{WP*}_{h,t} ) \nonumber \\
  & \cdot (a \sum_{l \in \mathcal{N}}(u_{l,t} - w_{l,t} - x^*_{l,t}) + b ) ] ) \nonumber \\
  & - ( \sum_{h \in \mathcal{N}^{WP}_n} \mathbb{E}[  (u_{h,t} - w_{h,t} -  x^{WP*}_{h,t} ) \nonumber \\
  & \cdot (a \sum_{l \in \mathcal{N}_n}(u_{l,t} - w_{l,t} - x^*_{l,t}) + b ) ]  \nonumber \\
& + \sum_{j \in \mathcal{N}^{LS}_n} \mathbb{E}[ (u_{j,t} - w_{j,t} -  x^{LS*}_{j,t} ) \nonumber \\
  & \cdot (a \sum_{l \in \mathcal{N}}(u_{l,t} - w_{l,t} - x^*_{l,t}) + b )  ] +   C_{B,t}(n_{\sigma},n) {x_{n,tot,t}^*} )  \nonumber \\
  = & R^{WP}_t \sum_{h \in \mathcal{N}^{WP}_n }x^{WP*}_{h,t} + R^{LS}_t \sum_{j \in \mathcal{N}^{LS}_n} {x^{LS*}_{j,t}}  \nonumber \\
  & - C_{B,t}(n_{\sigma},n) {x_{n,tot,t}^*} \nonumber \\
  \geq  & R^{WP}_t  {x^{WP*}_{n,tot,t}} +  R^{LS}_t {x^{LS*}_{n,tot,t}} - C_{B,t}(n_{\sigma},n) {x_{n,tot,t}^*}  \nonumber \\
  := &  \widehat{Z}_{n,t} (R^{WP}_t,R^{LS}_t,n_{\sigma}),
\end{align}
where
\begin{align}\label{51}
x^*_{l,t} = \left\{ \begin{array}{cc}
                            x^{WP*}_{l,t}, & \hbox{if $l \in \mathcal{N}^{WP}_n$,} \\
                            x^{LS*}_{l,t}, & \hbox{if $l \in \mathcal{N}^{LS}_n$.}
                          \end{array}
\right.
\end{align}
In the first equality in (\ref{l}), the first term (lines 1-3) represents the expected income from WP and LS prosumers and the second term (lines 4-7) represents the expected payment to DAM and EBM. The second equality holds by eliminating the common terms and using the sum of the EEE (\ref{eb}) of all LS prosumers, i.e.,
\begin{align}\label{}
& \sum_{j \in \mathcal{N}^{LS}_n} B_{j,t} -\sum_{j \in \mathcal{N}^{LS}_n} \mathbb{E}[(u_{j,t} - w_{j,t} - x^{LS*}_{j,t}) \nonumber \\
& \cdot (a \sum_{l \in \mathcal{N}_n}(u_{l,t} - w_{l,t} - x^{*}_{l,t}) + b )] = R^{LS}_t \sum_{j \in \mathcal{N}^{LS}_n} {x^{LS*}_{j,t}}.
\end{align}
The inequality in (\ref{l}) holds by using (\ref{e1}) and (\ref{e2}). Since $x^{WP*}_{n,tot,t}$, $x^{LS*}_{n,tot,t} $ and $x_{n,tot,t}^*$ are functions of $n$, the lower bound of the expected EA's net profit can be calculated by a weighted sum of $\widehat{Z}_{n,t}$ in terms of $Q(n)$, i.e.,
\begin{align}\label{52}
{I}_t & (R^{WP}_t,R^{LS}_t,n_{\sigma}) : =  \sum_{n \in \mathcal{N}^+} Q(n)\widehat{Z}_{n,t} (R^{WP}_t,R^{LS}_t,n_{\sigma}).
\end{align}
Note that in the expression of $I_t$, $n$, $R^{WP}_t$ and $R^{LS}_t$ are inexplicitly contained in $x^{WP*}_{n,tot,t}$, $x^{LS*}_{n,tot,t}$ and $x^{*}_{n,tot,t}$ (see (\ref{a2}), (\ref{e1}) and (\ref{e2})). To realize the budget recovery of EA, we consider the constraint $I_t(R^{WP}_t,R^{LS}_t,n_{\sigma}) \geq 0$, $\forall n_{\sigma} \in \mathcal{N}^{\pm}$, $\forall t \in \mathcal{T}$. \footnote[4]{As seen from (\ref{l}), $\widehat{Z}_{n,t}$ is the lower bound of the net profit of EA in any $n$-WP prosumer scenario, i.e., $Z_{n,t}$. Hence, we employ $\widehat{Z}_{n,t}$ in (\ref{52}) to produce a more conservative solution.}

\subsection{Two-level Stochastic Convex Programming Algorithm}

\begin{figure}[htbp]
  \centering
  \includegraphics[height=5cm,width=8cm]{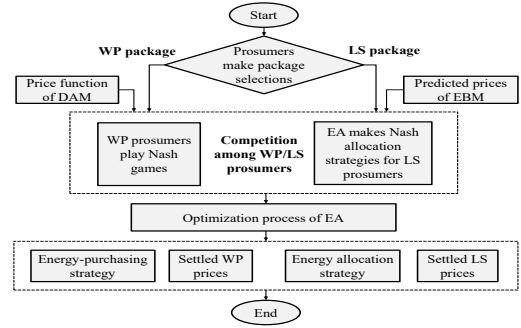}\\
  \caption{Overall decision-making process of WP/LS prosumers and EA.}\label{ff9}
\end{figure}

To facilitate the formulation of the constraint of the ramp rate limits, we define
\begin{align}\label{}
x^*_{0,tot,{t}}:= & \frac{Nb-N R_t^{LS}}{a(N+1)} +\sum_{i \in \mathcal{N}}(u_{i,t} - \mu_{i,t}), \\
 x^*_{N,tot,{t}} := & \frac{Nb-N R_t^{WP}}{a(N+1)} +\sum_{i \in \mathcal{N}}(u_{i,t} - \mu_{i,t}),
\end{align}
which are obtained by letting $n=0$ and $N$ in (\ref{a2}), respectively. Obviously, if $R_t^{WP} \geq R_t^{LS}$, $x^*_{0,tot,{t}}$ and $x^*_{N,tot,{t}}$ are the upper and lower bounds of $ x^*_{n,tot,t}$, respectively, and vice versa. Based on the discussion in {Section \ref{sb}}, with certain given $n_{\sigma} \in \mathcal{N}^{\pm}$, the expression of ${W}_{n,t}( R_t^{WP}, R_t^{LS} ; C_{B,t})$ is deterministic in each sub-space, $\forall n \in \mathcal{N}^+$. Therefore, in the proposed two-level stochastic convex programming algorithm, we firstly solve a social cost minimization problem in each sub-space, and then find the global optimal solution by choosing the minimum function value over all these sub-spaces. A basic optimization problem for any ${n}_{\sigma} \in \mathcal{N}^{\pm}$ and $t \in \mathcal{T}$ is formulated in Problem (P2).

 \begin{align}\label{uq}
 \hbox{\textbf{(P2)}}: \quad & \hbox{\em{Case 1:}} \nonumber \\
\min  \limits_{R_t^{WP},R_t^{LS}} & \sum_{{n} \in \mathcal{N}^+} Q(n) {W}_{n,t}(R_t^{WP},R_t^{LS}; C^{\geq}_{B,t}(n_{\sigma},n))\\
   s.t.  \quad & R_t^{WP} \geq R_t^{LS} , \tag{\theequation-a}\\
   & (R_t^{WP},R_t^{LS}) \in {{\mathcal{F}}_{{n}_{\sigma},t}}, \tag{\theequation-b}\\
& R_t^{WP} \geq \underline{R}_t^{WP}, \tag{\theequation-c}\\
    & R_t^{LS} \geq \underline{R}_t^{LS},
          \tag{\theequation-d}\\
     &  {{{I_t}}(R^{WP}_t,R^{LS}_t,{n}_{\sigma})} \geq 0, \tag{\theequation-e}\\
    &x^*_{0,tot,{t}} -x^*_{tot,t-1}  \leq \bar{x}_t^{lim}, \tag{\theequation-f}\\
    &x^*_{N,tot,{t}} -x^*_{tot,{t-1}} \geq \underline{x}_t^{lim}. \tag{\theequation-g}
    \end{align}
   \begin{align}\label{up}
& \hbox{\em{Case 2:}} \nonumber \\
\min  \limits_{R_t^{WP},R_t^{LS}} & \sum_{{n} \in \mathcal{N}^+} Q(n) {W}_{n,t}(R_t^{WP},R_t^{LS}; C^{<}_{B,t}(n_{\sigma},n))\\
   s.t.  \quad & R_t^{WP} < R_t^{LS} , \tag{\theequation-a} \\
   & (R_t^{WP},R_t^{LS}) \in \bar{{\mathcal{F}}}_{{n}_{\sigma},t}, \tag{\theequation-b}\\
& R_t^{WP} \geq \underline{R}_t^{WP}, \tag{\theequation-c}\\
    & R_t^{LS} \geq \underline{R}_t^{LS},
          \tag{\theequation-d}\\
   &  {{{I_t}}(R^{WP}_t,R^{LS}_t,{n}_{\sigma})} \geq 0, \tag{\theequation-e}\\
    &x^*_{N,tot,{t}} -x^*_{tot,t-1} \leq \bar{x}_t^{lim}, \tag{\theequation-f}\\
    &x^*_{0,tot,{t}} -x^*_{tot,{t-1}} \geq \underline{x}_t^{lim}. \tag{\theequation-g}
\end{align}
$x^*_{tot,{t-1}}$ denotes the settled quantity of total balancing energy at time $t-1$ and thus is known. (\ref{uq}-f), (\ref{uq}-g), (\ref{up}-f) and (\ref{up}-g) jointly define the ramp rate limits of the EBM with $\bar{x}_t^{lim} > 0 > \underline{x}_t^{lim}$, $t \in \mathcal{T}$. Essentially, we consider the upper (lower) bound of balancing energy in the constraint of ramp-up (ramp-down) rate limit to ensure a more conservative optimization result. The optimal solution, defined by $(R_{t,n_{\sigma}}^{WP*},R_{t,n_{\sigma}}^{LS*})$, of Problem (P2) is obtained by choosing the better solution between Cases 1 and 2.
The entire two-level stochastic convex programming algorithm is stated in Algorithm \ref{ths}.
The overall decision-making process of WP/LS prosumers and EA is shown in Fig. \ref{ff9}.

\begin{algorithm}
\caption{Two-level Stochastic Convex Programming Algorithm}\label{ths}
\begin{algorithmic}[1]
\State Initialize all relevant parameters of the market;
\For {$t=1,2,3,...,$}
\State {\em{Lower-Level Algorithm}} (steps 4-7)
\For {$n_{\sigma} = -1,0,1,...,N$,}
\State Solve {Problem (P2)};
\State Record inversion point $n_{\sigma}$ associated with the optimal solution $(R_{t,n_{\sigma}}^{WP*},R_{t,n_{\sigma}}^{LS*})$, and calculate $\sum_{{n} \in \mathcal{N}^+} Q(n) {W}_{n,t}(R_{t,n_{\sigma}}^{WP*},R_{t,n_{\sigma}}^{LS*} ; C_{B,t})$;
\EndFor;
\State {\em{Upper-Level Algorithm}} (steps 9-12)
\State Locate the sub-space of the global optimal solution by seeking
\begin{align}\label{}
n^{*}_{\sigma} =  & \arg \min_{n_{\sigma} \in \mathcal{N}^{\pm}} \sum_{{n} \in \mathcal{N}^+} Q(n) {W}_{n,t} (R_{t,n_{\sigma}}^{WP*}, R_{t,n_{\sigma}}^{LS*} ; C_{B,t});
\end{align}
\State Obtain the global optimal solution by
\begin{align}\label{47}
(R_{t,n^{*}_{\sigma}}^{WP*},R_{t,n^{*}_{\sigma}}^{LS*}) = (R_{t,n_{\sigma}}^{WP*}, R_{t,n_{\sigma}}^{LS*})_{n_{\sigma} = n^{*}_{\sigma}};
\end{align}
\State Calculate the optimal energy-purchasing/allocation strategy $(x^{WP*}_{h,t},x^{LS*}_{j,t})$ by
\begin{align}
& x^{WP*}_{h,t} =  u_{h,t} - \mu_{h,t} + \frac{b + \sum_{f \in \mathcal{N},f \neq h}R^*_{f,t} -NR_{t,n^{*}_{\sigma}}^{WP*}}{a(N+1)}, \label{36}\\
& x^{LS*}_{j,t} =  u_{j,t} - \mu_{j,t} + \frac{b + \sum_{g \in \mathcal{N},g \neq j}R^*_{g,t} -NR_{t,n^{*}_{\sigma}}^{LS*}}{a(N+1)},\label{36+1}
\end{align}
with $R^*_{i,t} = \left\{ \begin{array}{cc}
                            R_{t,n^{*}_{\sigma}}^{WP*}, & \hbox{if $i \in \mathcal{N}^{WP}$,} \\
                            R_{t,n^{*}_{\sigma}}^{LS*}, & \hbox{if $i \in \mathcal{N}^{LS}$,}
                          \end{array}
\right.$ $h \in \mathcal{N}^{WP}$, $j \in \mathcal{N}^{LS}$;
\State Calculate the LS price for prosumer $j' \in \mathcal{N}^{LS}$ by
\begin{align}\label{hht}
   B^*_{j',t}  = &  R_{t,n^{*}_{\sigma}}^{LS*} x^{LS*}_{j',t}  + \mathbb{E} [ (u_{j',t} - w_{j',t} - x^{LS*}_{j',t}) \nonumber \\
  & \cdot (a \sum_{l \in \mathcal{N}}(u_{l,t} - w_{l,t} - x^*_{l,t}) + b ) ],
\end{align}
where $x^*_{l,t}$ is determined by (\ref{51}), $l \in \mathcal{N}$.
\EndFor
\end{algorithmic}
\end{algorithm}



\begin{Remark}\label{r1}
In {Algorithm \ref{ths}}, $x^{WP*}_{h,t} $ and $x^{LS*}_{j,t}$ are NE solutions based on the really happened selection scenarios in PB period, $h \in \mathcal{N}^{WP}$, $j \in \mathcal{N}^{LS}$. Therefore, there are $2^N$ possible combinations of sets $\mathcal{N}^{WP}$ and $ \mathcal{N}^{LS}$ in total. As a result, the calculation of $B^*_{j',t}$ in (\ref{hht}) is based on EEE (\ref{eb}) with the settled energy-purchasing/allocation strategies and LS prices, i.e., (\ref{47})-(\ref{36+1}), which also varies with different selection scenarios, $j' \in \mathcal{N}^{LS}$.
\end{Remark}


\section{Supplementary Discussion}\label{cd}

\subsection{The Principles of EEE Based WP/LS Pricing Scheme}\label{v1}

As the main purpose of this work, we aim to explore a promising market model with heterogenous price packages, which is rarely discussed in the existing research works. To this end, we particularly focus on the EEE based WP/LS price-package selection problem and provide a feasible solution to optimize the market. We believe that one may also explore other pricing schemes instead of the EEE based scheme, which is an open question that can be explored in the future.

To ensure the high feasibility of the proposed scheme, some key principles we have deliberately considered in this work are summarized as follows.
\begin{itemize}
  \item Explainable. The final LS prices are decided by EEE, which includes the costs generated from DA energy and balancing service. This enables the LS prosumers to be aware of the componential costs if they have the enquiry.
  \item Economical. The proposed WP/LS pricing scheme is based on the NE of all the package buyers, which can reflect the self-centric nature of market participants and provide relatively economical solutions for all the participants.
  \item Sustainable. The expected net profit of EA is analytically derived, which can help to realize the budget recovery of EA and prolong the service of EA.
  \item Reliable. The ramp rate limits of EBM are considered based on the ``worst'' possible selection results to enhance the reliability of the power system.
\end{itemize}

\subsection{The Adaptivity of Package-Selection Based Scheme}
In the proposed price-package selection based market, all the prosumers are allowed to choose any package they prefer. The factors that may influence their decisions can be complicated, e.g., cost, convenience, reputation of LS service, etc. Compared with single-package scheme, e.g., only wholesale price scheme, the proposed two-package based solution can provide more flexibilities to meet the different preferences of prosumers. In fact, the commonly discussed single-package scheme can be a special case of the two-package scheme, where one of the packages is not competitive and abandoned completely. This is because the objective function of the EA is based on the probability distribution of their selections. In case that either of the two packages is not appealing and no prosumer would like to choose during certain interval, then the probability of this scenario will be gradually decreased to zero and the influence of the abandoned package will be automatically eliminated. Otherwise, the influence of the two packages will co-exist in the market. In this sense, the probability based optimization scheme can adapt to the circumstance of the market to better match the preferences of the prosumers.

\subsection{Community Efficiency}
\begin{figure}[htbp]
  \centering
  \includegraphics[height=5cm,width=8cm]{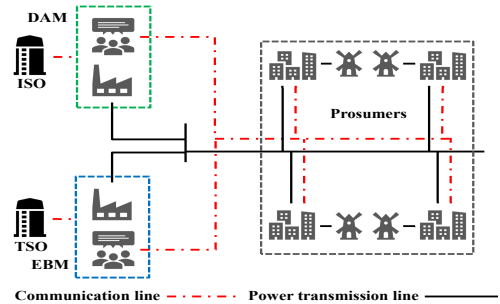}\\
  \caption{An illustrative framework of the USM.}\label{21}
\end{figure}

The performance of the proposed algorithm can also be evaluated by discussing the community efficiency (cost) compared with the corresponding uncoordinated single-package market (USM), e.g., only with wholesale price package and no EA is established. Specifically, we consider a USM \footnote[6]{The particularly concerned USM acts as a comparative case to verify the effectiveness of our optimization algorithm. We believe that in different comparative cases, the resulting social costs can be different, which is not to be discussed any further for conciseness.} where prosumers independently decide their energy-purchasing strategies with DAM and EBM by optimizing their own cost functions, i.e., play Nash games. In addition, the energy trading within the community is also allowed where certain prosumer sells its surplus energy to others (i.e., peer-to-peer fashion) with the up-regulation price\footnote[7]{The price of the energy provided by prosumers is designed to be equal to the up-regulation price since the energy buyers can pay a common unit price for the energy product from both prosumers and EBM, which creates a fair trading environment.} of balancing energy in the EBM. The framework of the discussed USM is illustrated in Fig. \ref{21}.

In this case, the incentive price of balancing energy for prosumer $i$ is $R_{i,t} = C_{UR,t}$ as shown in (\ref{uy}), which means the corresponding optimization problem of the prosumers can be formulated as
\begin{align}
   \textbf{(P1+)}: \quad & \min\limits_{x_{i,t}} \quad   {{U_{i,t}}(x_{i,t},\bm{x}_{-i,t}; C_{UR,t})}, \quad \forall i\in \mathcal{N},  \nonumber \\
    & s.t. \quad  \sum_{i \in \mathcal{N}}(u_{i,t}-\mu_{i,t}) - \sum_{i \in \mathcal{N}} x_{i,t} \geq 0.
\end{align}
Similar to Assumption \ref{a1}, we can assume $0 \leq b \leq \min \{C_{UR,t}| \forall t \in \mathcal{T}\}$. Then, analogous to the logic of the proof of Theorem \ref{l1}, the total demand of balancing energy of the community can be obtained as
\begin{align}\label{a2+1}
  & \hat{x}^*_{tot,t} := \frac{Nb - N C_{UR,t}}{a(N+1)} +\sum_{i \in \mathcal{N}}(u_{i,t}-\mu_{i,t}),
\end{align}
which is fixed since $C_{UR,t}$ is a systematic parameter. \footnote[8]{(\ref{a2+1}) can be directly obtained with (\ref{a2}) by letting $n=N$ and $R^{WP}_t=C_{UR,t}$, which means all the $N$ prosumers are with incentive price $C_{UR,t}$.}

Meanwhile, if EA only provides WP package, the total demand of balancing energy can be obtained by letting $n=N$ in (\ref{a2}), which gives
\begin{align}\label{a2+2}
  & x^*_{N,tot,t} = \frac{Nb - N R^{WP}_t}{a(N+1)} +\sum_{i \in \mathcal{N}}(u_{i,t}-\mu_{i,t}).
\end{align}
Note that (\ref{a2+2}) dominates (\ref{a2+1}) if the feasible region of $R^{WP}_t$ covers the fixed value $C_{UR,t}$, which means the social cost can be reduced by optimizing cost function (\ref{32}) with $R^{WP}_t$ ($R^{LS}_t$ is cancelled out, $Q(N)=1$, $Q(n)=0$, $\forall n \in \mathcal{N}^+\setminus \{N\}$).

When applying the two-package scheme, the social cost function is deduced as a weighted sum (\ref{32}), which is probability based compared with single-package counterpart. Then, there is no concrete comparison basis between single- and two-package schemes in terms of the social cost, since the package selection result of prosumers will definitely influence the final settled social cost. In this sense, our proposed two-package scheme emphasizes the convenience of prosumers by providing more choices, which is advantageous over the single-package scheme. Alternatively, if we simply aim to reduce the social cost of the community with single-package scheme, one can abandon either of the two packages, e.g., retain variable $R^{WP}_t$ and let $Q(N)=1$, $Q(n)=0$, $\forall n \in \mathcal{N}^+\setminus \{N\}$, $t\in \mathcal{T}$.

\subsection{Non-convexity of Stackelberg Game}\label{nc}

As discussed in Section \ref{sc}, Problem (P1) essentially is a nonlinear generalized NE problem. In Theorem \ref{l1}, Assumption \ref{a1} ensures an interior NE with constraint (\ref{co1}) inactivated (see the proof procedure of Theorem \ref{l1}). In case that the EA gives the pricing strategy of ${R}_t^{WP}$ and ${R}_t^{LS}$ in a wider range, e.g., $\underline{R}_t^{WP},\underline{R}_t^{LS} < b$, it is possible that the NE is nonlinearly determined by $(R^{WP}_t,R^{LS}_t)$  with constraint (\ref{co1}) activated. Then, the resulting social cost function can be non-convex in terms of $(R^{WP}_t,R^{LS}_t)$, which has been revealed by many works on Stackelberg games with nonlinear bi-level optimization frameworks \cite{cui2018two}. In this case, our proposed optimization scheme can provide an approximate optimal solution. To address the non-convexity issue of Stackelberg games, existing methods include branch-and-bound methods \cite{edmunds1991algorithms}, descent methods \cite{savard1994steepest}, penalty methods \cite{ishizuka1992double}, and trust region methods \cite{colson2005trust}, etc. Differently, in this work, we analytically derive the condition of $\underline{R}^{WP}_t$, $\underline{R}^{LS}_t$ and $b$, i.e., (\ref{sf}), which suffices to ensure an overall mixed integer quadratic programming problem with unique Stackelberg equilibrium and reduces the computational complexity (NP-complete \cite{del2017mixed}) compared with the corresponding nonlinear bi-level programming problem (NP-hard \cite{lv2008neural}).

\section{Simulation}\label{d}

\subsection{Simulation Setup}

\begin{figure}[htbp]
  \centering
  \includegraphics[height=7cm,width=8cm]{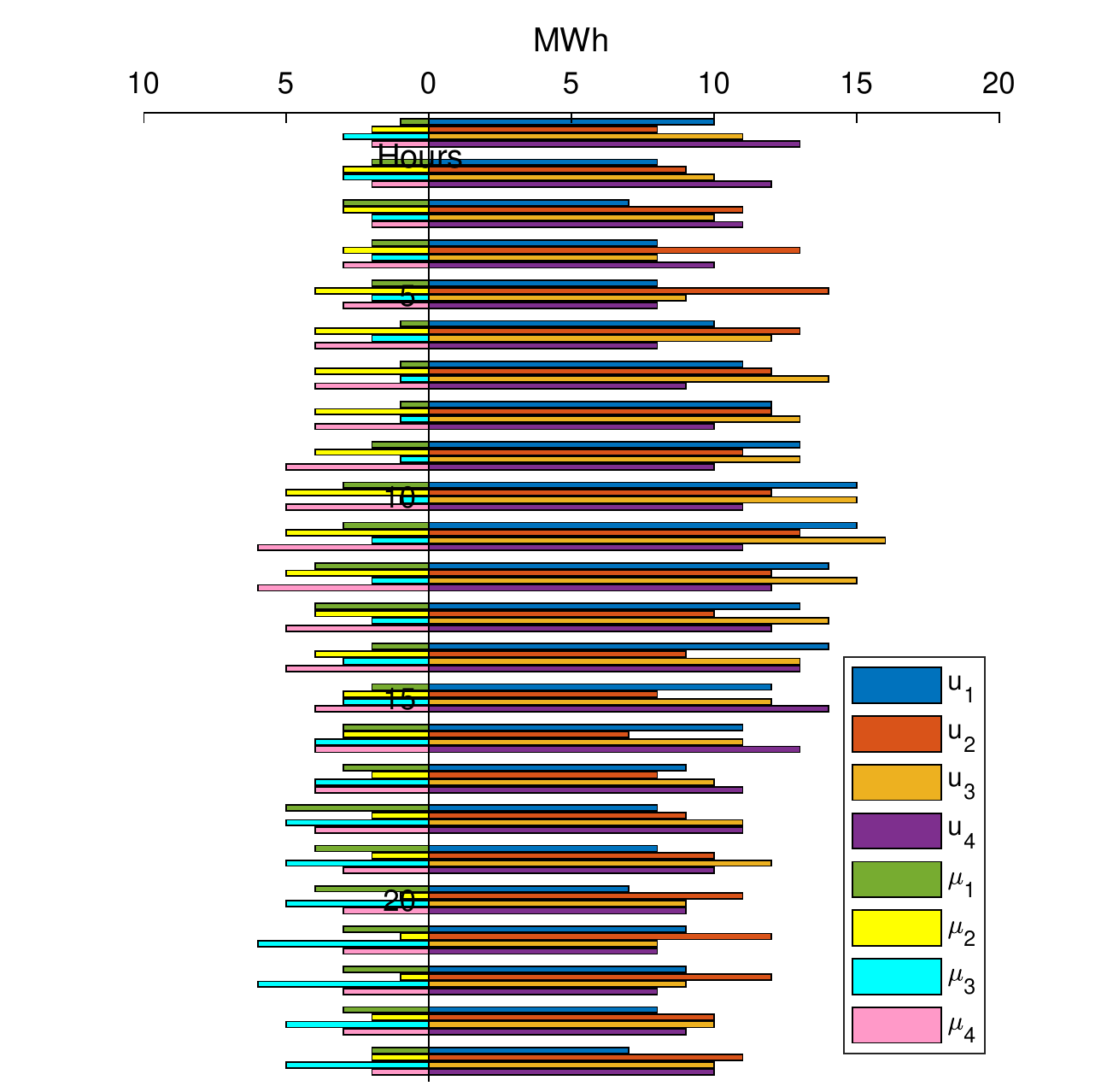}\\
  \caption{Demand of prosumers and mean output of WPPs in the next day.}\label{loadprofile2}
\end{figure}

\begin{figure}[htbp]
  \centering
  \includegraphics[height=3cm,width=8cm]{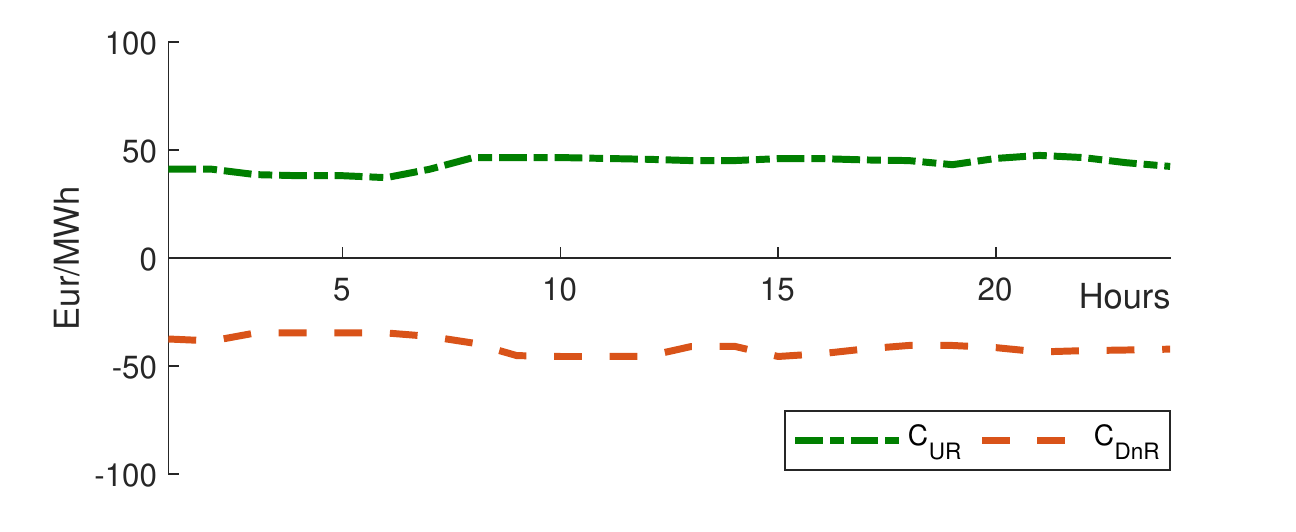}\\
  \caption{Up- and down- regulation price of FINGRID in Jun. 20th, 2018. (According to (\ref{sa1}), a negative down-regulation price means a positive charge to prosumers if they inject energy into EBM.)}\label{fin}
\end{figure}

In the simulation, we will testify the performance of our proposed optimization algorithm in a hourly optimization scheme during a day for a 4-prosumer community, e.g, $\mathcal{N}=\{1,2,...,4\}$ and $\mathcal{T}=\{1,2,...,24\}$. The parameters of the generation cost function of DAM are set as $a = 0.2$\texteuro $/$(MWh)$^{2}$, $b = 0.5$\texteuro $/$MWh, $c = 1$\texteuro. The demand of prosumers, $u_i:=(u_{i,1},...,u_{i,24})^T$, and the mean output of WPPs, $\mu_i:=(\mu_{i,1},...,\mu_{i,24})^T$, are shown in Fig. \ref{loadprofile2}, $\forall i \in \mathcal{N}$. The hourly regulation prices of the EBM, $C_{UR}:=(C_{UR,1},...,C_{UR,24})^T$ and $C_{DnR}:=(C_{DnR,1},...,C_{DnR,24})^T$, in the next day are shown in Fig. \ref{fin}, which are cited from the Finland electricity market (Jun. 20th, 2018) \cite{fin}. In our proposed DA optimization scheme, the regulation prices can be obtained by using some mature prediction techniques with low prediction errors, e.g., artificial neural networks \cite{sahay2013day}. In this simulation, we directly use the price data without any modification. For the WPPs, we let $\nu_{i,t}=20$ and $w^{ca}_i = 10$ MW, $\forall i \in \mathcal{N}$, $\forall t \in \mathcal{T}$. The parameters of Poisson binomial distribution of package selections are set as $(q_1,q_2,q_3,q_4)=(0.35,0.5,0.65,0.7)$. In addition, we let
$\bar{x}_t^{lim} =10$WM/h, $\underline{{x}}_t^{lim} = -10$WM/h, $\underline{R}_t^{WP} = \underline{R}_t^{LS} = $10 \texteuro$/$MWh, $\forall t \in \mathcal{T}$.\footnote[9]{In practice, the unit of the ramp rate limits may be given as minute-based depending on the specific power infrastructures. In this case, one just need to convert the hourly optimizations to minutely optimizations.} To shed light on the detailed selection result in PB period, we list all the possible package selection scenarios of prosumers in Table \ref{t1+11}. Obviously, in a 4-prosumer market, there are $2^4=16$ possible selection scenarios in total.


\begin{table}[htbp]
\caption{Combinations of WP/LS selections in a 4-prosumer market}
\label{t1+11}
\begin{center}
\begin{tabular}{ccccc}
\hline
 Scenario No. & Pros. 1 & Pros. 2 & Pros. 3 & Pros. 4\\
\hline
 1 & $\square$& $\square$ & $\square$ & $\square$ \\
 \hline
 2 &$\blacksquare$ & $\square$& $\square$& $\square$\\
 \hline
 3 &$\square$  &$\blacksquare$  &$\square$ &$\square$ \\
 \hline
 4 &$\square$ &$\square$ & $\blacksquare$& $\square$\\
 \hline
 5 & $\square$& $\square$ & $\square$ &$\blacksquare$\\
 \hline
 6 &$\blacksquare$ &$\blacksquare$&$\square$& $\square$  \\
 \hline
 7 & $\blacksquare$ &$\square$& $\blacksquare$&$\square$ \\
  \hline
 8 &$\blacksquare$ &$\square$ & $\square$&$\blacksquare$ \\
  \hline
 9 & $\square$ & $\blacksquare$ & $\blacksquare$& $\square$ \\
 \hline
 10 &$\square$ & $\blacksquare$& $\square$& $\blacksquare$\\
 \hline
 11 &$\square$ &$\square$ &$\blacksquare$ &$\blacksquare$ \\
 \hline
 12 &$\blacksquare$ &$\blacksquare$ & $\blacksquare$& $\square$\\
 \hline
 13 & $\blacksquare$& $\blacksquare$&$\square$ &$\blacksquare$\\
 \hline
 14 &$\blacksquare$ &$\square$ &$\blacksquare$ &$\blacksquare$\\
 \hline
 15 & $\square$&$\blacksquare$ & $\blacksquare$&$\blacksquare$ \\
  \hline
 16 &$\blacksquare$ &$\blacksquare$ & $\blacksquare$&$\blacksquare$  \\
\hline
 \multicolumn{5}{l}{Note: $\square$: WP package; $\blacksquare$: LS package.}\\
 \hline
\end{tabular}
\end{center}
\end{table}

\begin{table}[htbp]
\caption{Settled $(R_{t,n^{*}_{\sigma}}^{WP*},R_{t,n^{*}_{\sigma}}^{LS*})$ in the next day.}\label{t2}
\begin{center}
\begin{tabular}{cccccc}
\hline
$t$ & $R _{t,n^{*}_{\sigma}}^{WP*}$& $R _{t,n^{*}_{\sigma}}^{LS*}$& $t$ & $R _{t,n^{*}_{\sigma}}^{WP*}$ & $R _{t,n^{*}_{\sigma}}^{LS*}$ \\
\hline
1 & 45.19 & 31.10 & 13 & 50.20 & 31.08 \\
 \hline
2 & 45.18 & 31.10 & 14 & 50.32 & 31.38 \\
 \hline
3 & 42.23 & 30.73 & 15 & 51.42 & 31.24 \\
 \hline
4 & 41.45 & 30.46 & 16 & 51.54 & 31.57 \\
 \hline
5 & 41.53 & 30.51 & 17 & 50.57 & 31.02 \\
 \hline
6 & 40.50 & 30.28 & 18 & 50.52 & 31.93 \\
 \hline
7 & 45.29 & 31.23 & 19 & 48.01 & 31.64 \\
 \hline
8 & 51.98 & 30.77 & 20 & 51.79 & 31.87 \\
 \hline
9 & 52.05 & 31.14 & 21 & 53.67 & 32.03 \\
 \hline
10 & 51.90 & 30.77 & 22 & 52.30 & 31.83 \\
 \hline
11 & 51.53 & 31.22 & 23 & 49.04 & 31.44 \\
 \hline
12 & 50.91 & 30.97 & 24 & 46.89 & 31.39 \\
\hline
\multicolumn{6}{l}{Unit of $R _{t,n^{*}_{\sigma}}^{WP*}$ and $R _{t,n^{*}_{\sigma}}^{LS*}$: \texteuro/MWh.} \\
\hline
\end{tabular}
\end{center}
\end{table}

\subsection{Simulation Result}
By {Algorithm 1}, the optimal pricing strategies, $(R _{t,n^{*}_{\sigma}}^{WP*},R_{t,n^{*}_{\sigma}}^{LS*})$, $\forall t \in \mathcal{T}$, are obtained in Table \ref{t2}. It can be seen that, with the given parameters, choosing LS package is more economical since $R_{t,n^{*}_{\sigma}}^{LS*}$ is smaller than $R _{t,n^{*}_{\sigma}}^{WP*}$ at all times. This phenomenon is on a case-by-case basis depending on the parameters in the constraints and Poisson distribution. Based on this result and the advantages of LS prices discussed in Sections \ref{s1} and \ref{cd}, the attraction of the LS price package is straightforward.

\begin{figure}[htbp]
  \centering
  \includegraphics[height=5cm,width=8cm]{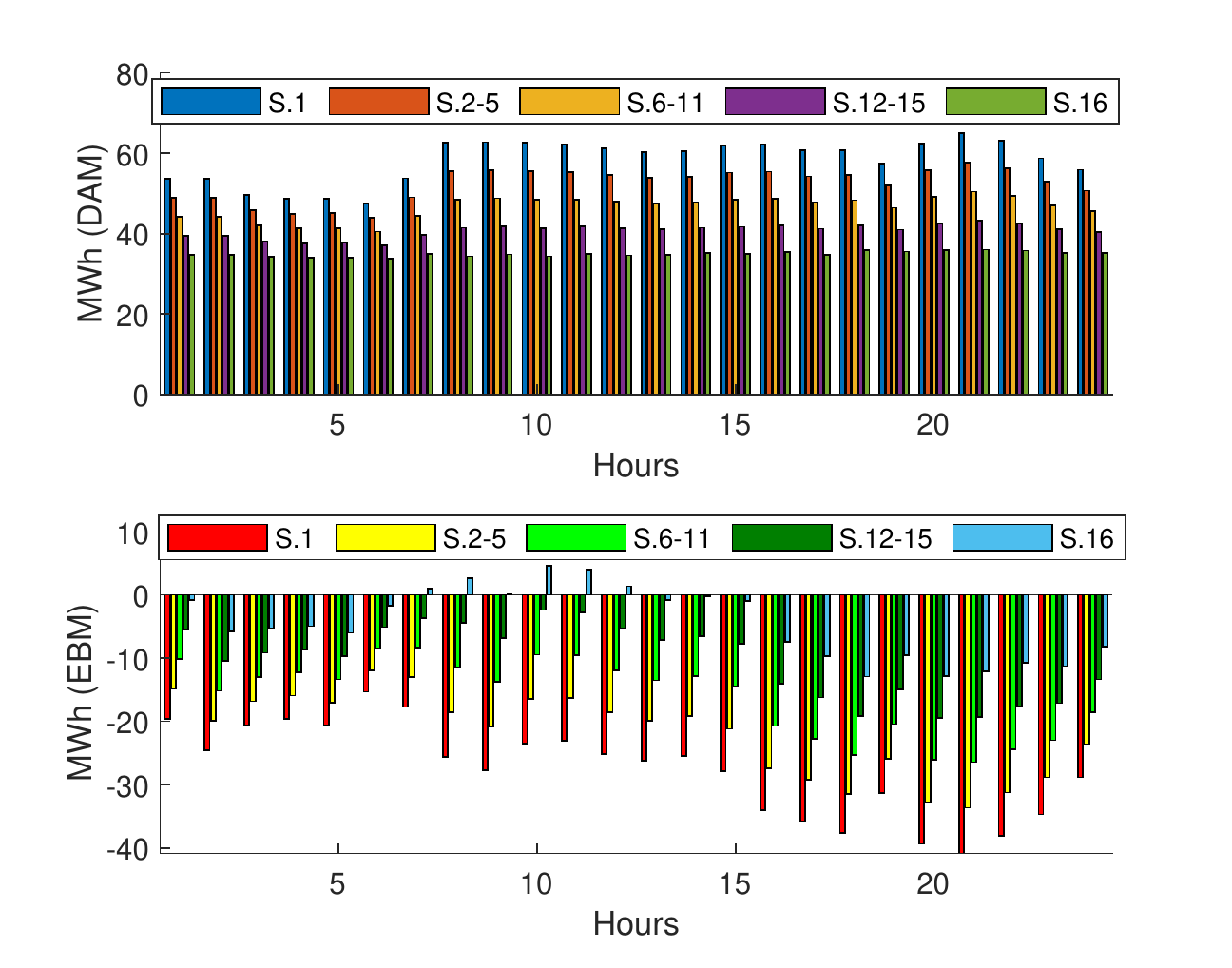}\\
  \caption{Energy demand of the community from DAM and EBM in the next day. (``S.'' is the abbreviation of ``scenario''.)}\label{de1}
\end{figure}

Fig. \ref{de1} shows how the total demand of the community is fulfilled by DAM and EBM in the next day. It worths noticing that, in any hour, the demand of balancing energy (also DA energy) in scenarios 2-5 (also scenarios 6-11 and 12-15 respectively) is consistent. This is because the number of WP prosumers is the same (i.e., 3 as shown in Table \ref{t1+11}) in all these scenarios, which results in the identical total demand of balancing energy as indicated by (\ref{a2}).

\begin{figure}[htpb]
  \centering
  \includegraphics[height=4cm,width=8cm]{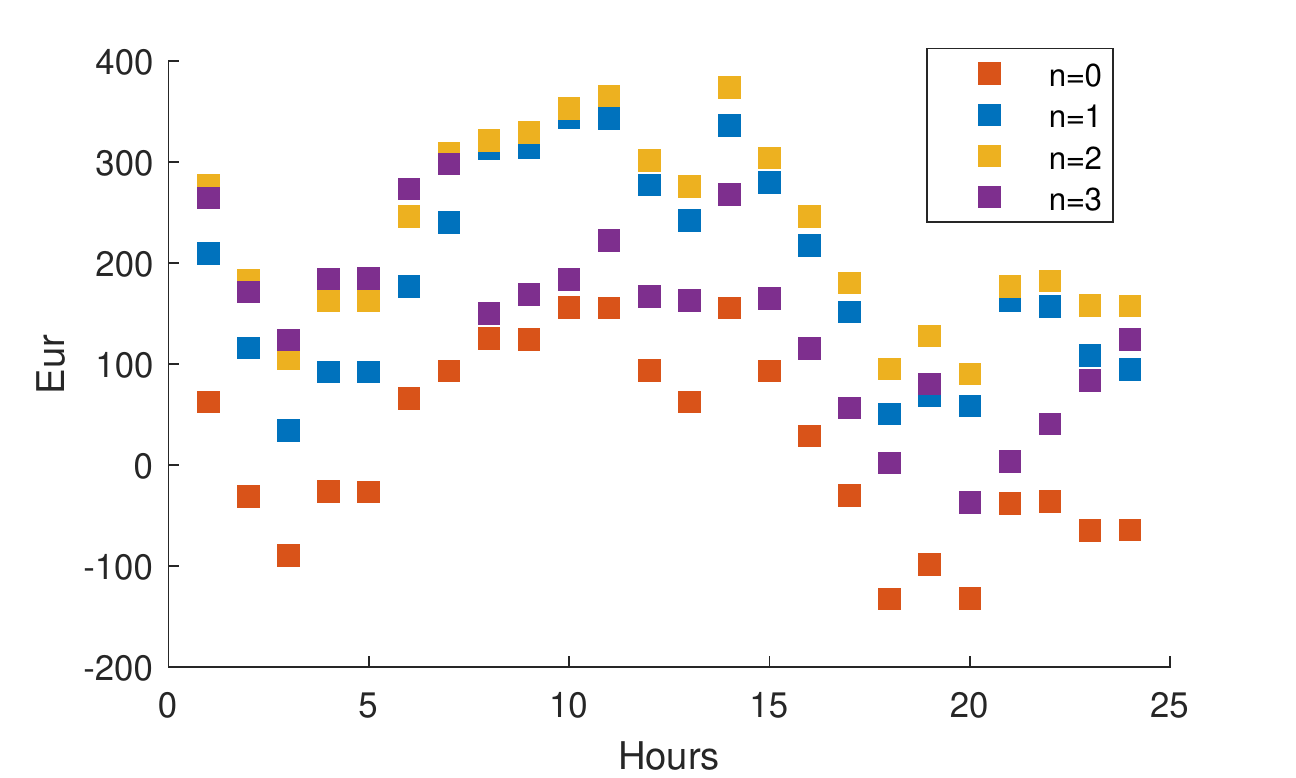}\\
  \caption{LS prices of prosumer 1 in the next day with $n=0,1,2,3$. (A negative LS price means a positive reward from EA.)}\label{lsprice}
\end{figure}

To show the determination of LS prices under different selection results in PB period (see Remark \ref{r1} for more explanations), we assume that prosumer 1 selects LS package and the number of total LS prosumers varies from 1 to 4 (i.e., $n=3,2,1,0$). By formula (\ref{hht}) and the settled $(R_{t,n^{*}_{\sigma}}^{WP*}, R _{t,n^{*}_{\sigma}}^{LS*})$ in Table \ref{t2}, the settled LS price $B^*_{1,t}$ in different hours is obtained as Fig. \ref{lsprice}, $\forall t \in \mathcal{T}$.

\begin{figure}[htpb]
  \centering
  \includegraphics[height=3cm,width=8cm]{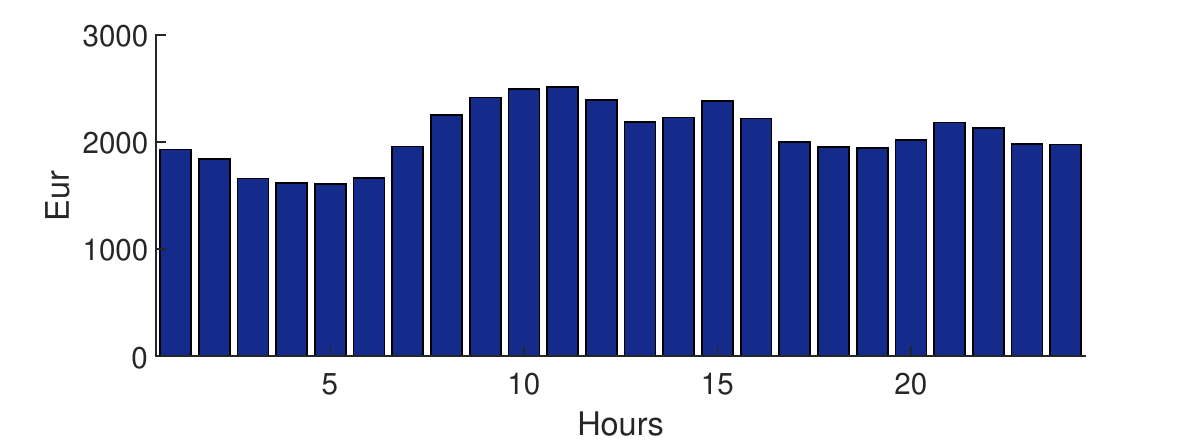}\\
  \caption{The expectation of the social cost in the next day.}\label{costrange}
\end{figure}

Fig. \ref{costrange} shows the expectation of the social cost in different hours in the next day, which is roughly within [\texteuro1600,\texteuro2500].

\begin{figure}[htbp]
\centering
  \includegraphics[height=5cm,width=8cm]{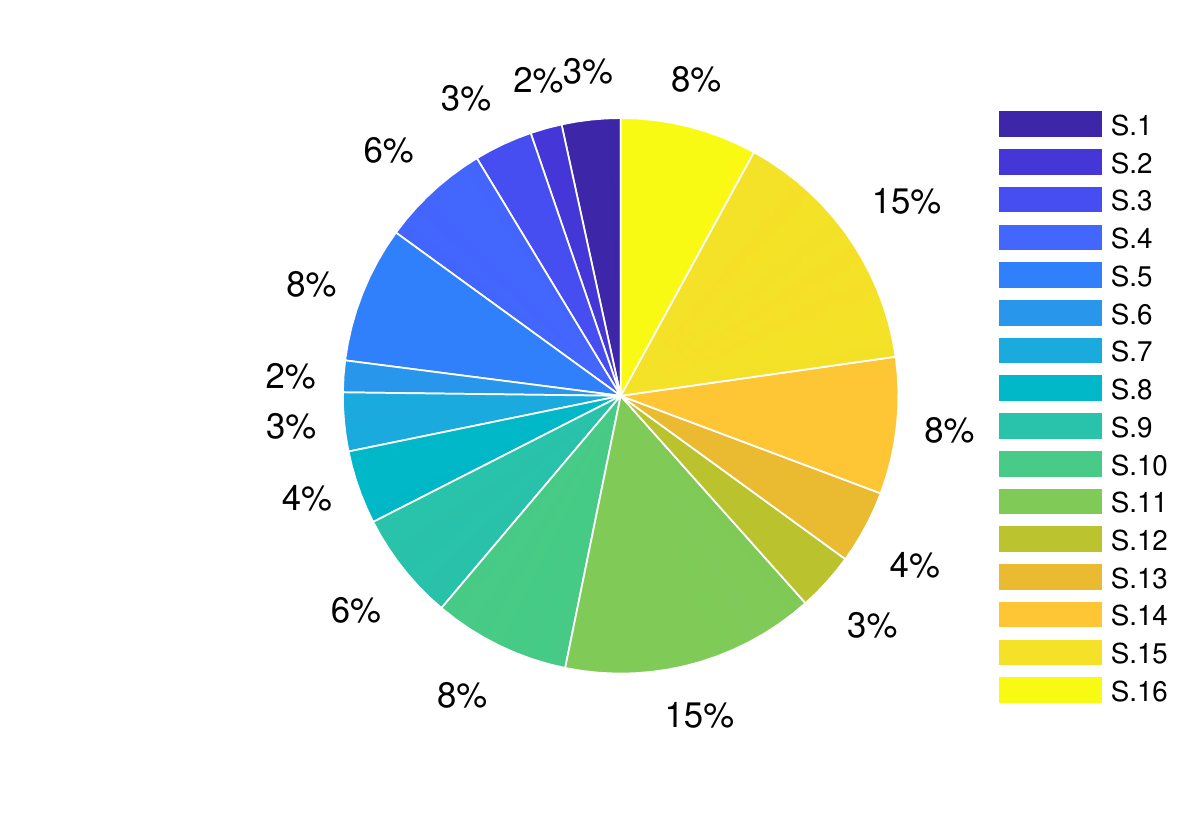}\\
  \caption{Probability distribution of the 16 package-selection scenarios during 4:00$\sim$5:00 AM in the next day.}\label{weightsumpoa1}
\end{figure}

\begin{figure}[htbp]
\centering
  \includegraphics[height=3cm,width=8cm]{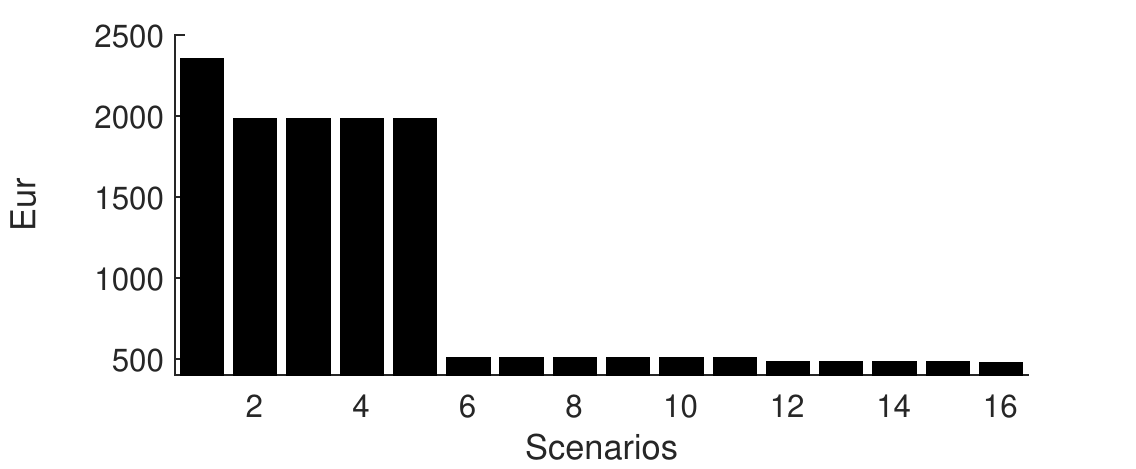}\\
  \caption{Social cost distribution of the 16 package-selection scenarios during 4:00$\sim$5:00 AM in the next day.}\label{weightsumpoa2}
\end{figure}

In addition, the social cost can be analyzed by depicting the probability and cost distributions in different hours. For the clearness, we take the 5th hour (i.e., 4:00$\sim$5:00 AM) as an example. The probability distribution of the 16 package-selection scenarios is depicted in Fig. \ref{weightsumpoa1} and the corresponding social cost distribution is shown in Fig. \ref{weightsumpoa2}. It can be seen that the scenarios that most likely to happen are scenarios 11 and 15 with probability around $15\%$; the scenarios with the least probability are scenarios 2 and 6 with probability around $2\%$. Meanwhile, the minimum and maximum social costs are located in scenario 16 (around \texteuro 475) and scenario 1 (around \texteuro 2350), respectively, which means that the community has to reserve at least around \texteuro 2350 for the energy consumption within 4:00$\sim$5:00 AM according to our proposed optimization algorithm.

\section{Conclusion}\label{s5}

In this paper, we proposed an economical pricing strategy for EA to optimize the social cost of a prosumer community in a future two-settlement based electricity market. To meet the different preferences of the prosumers, WP and LS packages are provided by the EA. Analytical NE solutions of WP and LS prosumers were derived based on the proposed PB-QB mechanism. We employed Poisson binomial distribution to model the package selections of prosumers, and finally a stochastic Stackelberg game was formulated between the prosumers and EA. A two-level stochastic convex programming algorithm was proposed for the EA to minimize the expectation of the social cost by considering the budget recovery of EA and the ramp rate limits of EBM.
In the proposed package-selection scheme, we considered several beneficial principles, e.g., explainable, economical, sustainable and reliable. In the future works, it is promising to explore more market models and multi-package based incentive methods for the market manager to improve the social benefit of the community.

\appendix{}

\subsection{Proof of Theorem \ref{l1}}\label{l1p}

By (\ref{uy}), ${U_{i,t}}(x_{i,t},  \bm{x}_{-i,t} ;R_{i,t})$ is smooth and convex at $x_{i,t}$, $\forall i \in \mathcal{N}$. We firstly solve an unconstrained optimization problem for the prosumers, where the best response of prosumer $i$, $x_{i,t}^*$, is obtained by solving the following first-order optimality condition
\begin{align}\label{eed11}
  x_{i,t}^{*} & =  \arg \min\limits_{x_{i,t}} {U_{i,t}} (x_{i,t},  \bm{x}_{-i,t};  R_{i,t}) \nonumber \\
   & =  \arg ( \nabla_{x_{i,t}}U_{i,t} (x_{i,t},  \bm{x}_{-i,t} ;R_{i,t})=0),
\end{align}
which gives
\begin{align}\label{eed}
2x^*_{i,t} + & \sum_{z \in \mathcal{N} \setminus \{i\}}x_{z,t} = 2 u_{i,t} - 2 \mu_{i,t} \nonumber \\
 &  + \frac{b-{R}_{i,t}}{a} + \sum_{z \in \mathcal{N}\setminus \{i\}}(u_{z,t}- \mu_{z,t}),
\end{align}
with $R_{i,t}$ determined by (\ref{15}). By expanding (\ref{eed}) to the whole prosumer set $\mathcal{N}$, a compact equation
$\bm{A}\bm{x}_t^*=\bm{B}_t$ can be obtained, where
\begin{align}\label{}
\bm{A}=\left[
         \begin{array}{cccc}
           2 &  1 & \cdots    & 1 \\
            1 &  2 & \cdots     & 1 \\
            \vdots & \vdots &  \ddots &   \vdots \\
           1 & 1 & \cdots &  2 \\
         \end{array}
       \right] \in \mathbb{R}^{N \times N}, \quad \bm{B}_t= \nonumber
\end{align}
\begin{align}\label{}
\left[
  \begin{array}{c}
    2 u_{1,t}  - 2 \mu_{1,t}+ \sum_{z \in \mathcal{N}\setminus \{1\}}(u_{z,t}- \mu_{z,t}) + \dfrac{b-{R}_{1,t}}{a} \\
    \vdots \\
    2 u_{i,t}  - 2 \mu_{i,t}+ \sum_{z \in \mathcal{N}\setminus \{i\}}(u_{z,t}- \mu_{z,t}) + \dfrac{b-{R}_{i,t}}{a} \\
    \vdots \\
    2 u_{N,t}  - 2 \mu_{N,t}+ \sum_{z \in \mathcal{N}\setminus \{N\}}(u_{z,t}- \mu_{z,t}) + \dfrac{b-{R}_{N,t}}{a} \\
  \end{array}
\right]. \nonumber
\end{align}
Since $\bm{A}$ is invertible, we can have
\begin{equation}\label{}
  \bm{x}_t^*=\bm{A}^{-1}\bm{B}_t = \left[\begin{array}{ccc}
                       \dfrac{N}{N+1} & \cdots & \dfrac{-1}{N+1}  \\
                       \vdots & \ddots &  \vdots \\
                       \dfrac{-1}{N+1} &  \cdots & \dfrac{N}{N+1}
                     \end{array}
\right] \bm{B}_t,
\end{equation}
which gives
\begin{align}\label{ar1}
x^{*}_{i,t} =  u_{i,t} - \mu_{i,t} + \frac{b + \sum_{z \in \mathcal{N}, z \neq i}R_{z,t} -N{R}_{i,t}}{a(N+1)}.
\end{align}
Therefore, in an $n$-WP prosumer community, the total demand of balancing energy can be obtained by\footnote[5]{Operation $\sum_{i \in \mathcal{E}; \mathcal{A}} e_i$ means summing up scalar $e_i$ over set $\mathcal{E}$ subject to condition $\mathcal{A}$.}
\begin{align}
    \sum_{i \in \mathcal{N}; |\mathcal{N}^{WP}| = n} x^{*}_{i,t}   = & \frac{Nb - (n R_t^{WP} + (N-n) R_t^{LS})}{a(N+1)}   \nonumber \\
   &  +\sum_{i \in \mathcal{N}}(u_{i,t}-\mu_{i,t}).
\end{align}
Then, by Assumption \ref{a1}, we can have
\begin{align}\label{75}
g_t & (\bm{x}^*_{t}) =  \sum_{i \in \mathcal{N}}(u_{i,t}-\mu_{i,t}) -  \sum_{i \in \mathcal{N}; |\mathcal{N}^{WP}| = n} x^{*}_{i,t}  \nonumber \\
 = & \frac{(n R_t^{WP} + (N-n) R_t^{LS}) -Nb}{a(N+1)} \geq 0.
\end{align}
By (\ref{eed11}) and (\ref{75}), it can be checked that $(\bm{x}^*_t,0)$ (i.e., $\lambda_t^*=0$) is the solution to (\ref{b1}) and (\ref{b2}), which means $\bm{x}^*_t$ is an NE of Problem (P1). Hence, (\ref{alr1}) and (\ref{alr2}) can be obtained by letting ${R}_{i,t}={R}^{WP}_{t}$ if $i \in \mathcal{N}^{WP}$ and ${R}_{i,t}={R}^{LS}_{t}$ if $i \in \mathcal{N}^{LS}$, respectively. The total demand of the community can be obtained as (\ref{a2}) since
\begin{align}\label{}
x^*_{n,tot,t} = \sum_{i \in \mathcal{N}; |\mathcal{N}^{WP}| = n} x^{*}_{i,t}, \quad n \in \mathcal{N}^+, t \in \mathcal{T}.
\end{align}
The proof is completed.

\bibliographystyle{IEEEtran}

\bibliography{myref}

\end{document}